\documentclass[12pt,english,preprint]{aastex}
\usepackage[latin1]{inputenc}
\makeatletter

\shorttitle{Modeling Type I X-ray bursts with the Low Mach Number Approximation}
\shortauthors{D.J. Lin et al.}

\newcommand{\ddt}[1]{\frac{\partial #1} {\partial t}}

\newcommand{\ddz}[1]{\frac{\partial #1} {\partial z}}
\renewcommand{\vec}[1]{\mbox{\boldmath $ #1 $}}
\newcommand{\vel}{\vec{v}}
\newcommand{\gvec}{\vec{g}}
\newcommand{\grad}{\nabla}
\newcommand{\diver}{\nabla \cdot}
\newcommand{\ihat}{\vec{\hat{i}}}
\newcommand{\jhat}{\vec{\hat{j}}}
\newcommand{\khat}{\vec{\hat{k}}}
\newcommand{\dd}[2]{\frac{\partial #1}{\partial #2}}
\newcommand{\ddd}[2]{\frac{\partial^2 #1}{\partial #2 ^2}}
\newcommand{\matder}[1]{\frac{D #1}{Dt}}
\newcommand{\ijk}{(i,j,k)}

\newcommand{\n}{(n)}
\newcommand{\npo}{(n+1)}
\newcommand{\power}{_{\ijk}^{\n}}
\newcommand{\pluspower}{_{\ijk}^{\npo}}

\newcommand{\rhoprime}{\rho^{\prime}}

\makeatother

\usepackage{babel}
\makeatother
\begin{document}

\title{Low Mach number modeling of Type I X-ray burst deflagrations}

\author{David J. Lin}

\affil{Northwestern University, Department of Physics and Astronomy, 2145
Sheridan Road, Evanston, IL 60208, USA}

\email{d-lin@northwestern.edu}

\author{Alvin Bayliss}

\affil{Northwestern University, Department of Engineering Sciences and Applied
Mathematics, 2145 Sheridan Road, Evanston, IL 60208, USA}

\email{a-bayliss@northwestern.edu}

\and{}

\author{Ronald E. Taam}

\affil{Northwestern University, Department of Physics and Astronomy, 2145
Sheridan Road, Evanston, IL 60208, USA}

\email{r-taam@northwestern.edu}

\begin{abstract}
The Low Mach Number Approximation (LMNA) is applied to 2D hydrodynamical
modeling of Type I X-ray bursts on a rectangular patch on the surface
of a non-rotating neutron star. Because such phenomena involve decidedly
subsonic flows, the timestep increase offered by the LMNA makes routine
simulations of these deflagrations feasible in an environment where
strong gravity produces significant stratification, while allowing
for potentially significant lateral differences in temperature and
density. The model is employed to simulate the heating, peak, and
initial cooling stages in the deep envelope layers of a burst. During
the deflagration, Bénard-like cells naturally fill up a vertically
expanding convective layer. The Mach number is always less than 0.15
throughout the simulation, thus justifying the low Mach number approximation.
While the convective layer is superadiabatic on average, significant
fluctuations in adiabaticity occur within it on subconvective timescales.
Due to convective layer expansion, significant compositional mixing
naturally occurs, but tracer particle penetration through the convective
layer boundaries on convective timescales is temporary and spatially
limited. Thus, mixing occurs on the relatively slow burst timescale
through thermal expansion of the convective layer rather than from
mass penetration of the convective layer boundary through particle
convection. At the convective layer boundaries where mixing is less
efficient, the actual temperature gradient more closely follows the
Ledoux criteria. 
\end{abstract}


\section{Introduction}

In the thermonuclear flash model, Type I X-ray bursts (henceforth
referred to as \textit{bursts}) are understood to be caused by the
explosive ignition of hydrogen and/or helium gas, which have accreted
onto the outer surface of neutron stars from relatively low-mass,
binary companion donors. Extensive theoretical calculations using
diffusional-thermal and 1D hydrodynamical models have successfully
reproduced many of the general observational features of these bursts,
such as the energies involved ($\sim$10$^{38}$-10$^{39}$ ergs),
their rise times (seconds), durations ($\sim$ 10-100 seconds), spectral
softening, and recurrence intervals (several hours). (For reviews,
see Taam 1985, Lewin et al. 1995, Cumming 2004, and Strohmayer \&
Bildsten 2006.) However, multidimensional hydrodynamic modeling of
bursts has been much more limited, partly due to limitations in computational
resources. Using FLASH (Fryxell et al. 2000) based on the Piecewise
Parabolic Method (PPM) (Colella \& Woodward 1984), Zingale et al.
(2001) simulated bursts as 2D detonations, albeit assuming very low
mass accretion rates. Also using FLASH, Zingale et al. (2002) simulated
bursts as 2D deflagrations using an artificial temperature perturbation
in lieu of a realistic burning network. Spitkovsky et al. (2002) used
the shallow water approximation to examine how flames would propagate
during bursts around a two-layer neutron star surface, but their method
is incompressible, assumes an ideal gas law, and does not account
for thermal diffusion. In a different context, Dearborn et al. (2006)
studied the related helium flash problem in the 3D cores of evolved
giant stars using Djehuty (Bazán 2003), which is based on an explicit
Lagrange-Eulerian hydrodynamic method.

Our current modeling effort simulates the heating, peak, and initial
cooling stages of the deep envelope layers undergoing a burst with
2D hydrodynamics using the Low Mach Number Approximation (LMNA) and
remedies the shortcomings of previous work in several respects: (1)
compressibility effects are included; (2) potentially significant
lateral fluctuations in temperature and density are allowed; (3) the
essential input physics are considered, including thermal diffusion,
a realistic equation of state, and a burning network (a $3\alpha$
nuclear burning network is reported here for simplicity, and more
complete networks can be included); (4) no assumption is made regarding
the nature of convection; (5) sound waves are naturally excluded from
the domain, thereby eliminating both the acoustic timestep restriction
and acoustic boundary reflection problems which can plague fully compressible
simulations; and, (6) the computational time is reduced by a factor
of 10-100 compared to fully compressible methods, due to the corresponding
increase in the timestep.

Other approximation methods which eliminate the acoustic timestep
restriction include the Boussinesq approximation (Spiegel \& Veronis
1959; Miralles 2000), the anelastic approximation (Ogura \& Phillips
1962; Glatzmaier 1984), and implicit methods (e.g., Deupree 2000).
However, at present only the LMNA can successfully model the compressibility
effects and significant lateral fluctuations in temperature and density
which characterize deflagrations, as well as offer substantial increases
in timestep and avoid acoustic boundary complications. While the LMNA
is routinely used to model terrestrial combustion (e.g., Bayliss et
al. 1992; McGrattan et al. 2004), only recently have efforts begun
to adapt it to the astrophysical setting. Alternative astrophysical
LMNA models have been recently developed by Bell et al. (2004a) and
Almgren et al. (2006a, 2006b). Thus far, they have been used to examine
2D Landau-Darrieus planar flame instabilities during the early development
of a Type Ia supernova (Bell et al. 2004b), Rayleigh-Taylor unstable
flames in 2D and 3D (Bell et al. 2004c; Zingale 2005), and tested
against the anelastic approximation and other fully compressible methods
in a regime where all are valid (Almgren et al. 2006a, 2006b). The
model by Bell et al. neglects background stratification, because the
domain on which it was applied was much smaller than one pressure
scale height. The model by Almgren et al. does include background
stratification, and is thus more comparable to ours. However, our
LMNA model differs from Almgren et al.'s in several important respects:
(1) they evolve the density via the continuity equation, whereas we
evolve the temperature via the energy equation; (2) they neglect thermal
diffusion, because it is expected to be unimportant in their modeling
of Type Ia supernova, whereas we include it in our model; (3) they
allow a time-dependent background state, while presently, we assume
it is time-independent, a condition which is easily relaxed and is
reserved for future work; (4) we reformulate (using a novel function
which will be described in detail below) the perturbative pressure
gradient and buoyancy forces to accurately calculate the net vertical
force in the presence of significant cancellation for large gravity.

In $\mbox{\S}$2, the motivation and essence of our LMNA model are
presented. The results of applying the LMNA to simulate a Type I X-ray
burst in 2D are described in $\S$3. Next, $\S$4 briefly describes
verification studies which were performed. Finally, a discussion of
the key findings and limitations of the present model, along with
ideas for future development and applications, is presented in $\S$5.

\section{Computational procedure}

Weakly compressible (low Mach number) models have been extensively
employed in studies of terrestrial combustion. Such models are suitable
for low speed deflagrations. They account for thermal expansion of
the gas, but assume only weak compressibility. As such they allow
a filtering of sound waves from the hydrodynamic equations so that
timesteps employed in computational methods can be based on the local
fluid velocity rather than the (much larger) sound speed.

The essence of such models is to expand the hydrodynamic quantities
around a mean state with perturbations proportional to $M^{2}$, where
$M$ is the maximum local Mach number:

\[
A\simeq A^{0}+M^{2}A^{\prime}.\]
 Here, $A$ is the non-dimensionalized form of any of the hydrodynamic
quantities (density $\rho$, velocity $\vel$, pressure $P$, temperature
$T$, and mass fractions $X_{l}$ of species $l$ in the gas). This
expansion is then incorporated into the hydrodynamical equations (either
reacting Euler or reacting Navier-Stokes equations, together with
the equation of state). In terrestrial combustion, the zeroth order
pressure $P^{0}$ can be shown to be independent of the spatial coordinates,
and for an open system, $P^{0}$ is independent of time as well. Under
the low Mach number assumption, the pressure perturbation $P^{\prime}$
appears only in the momentum equation, and the pressure $P$ is assumed
to be equal to the constant $P^{0}$ in the energy and state equations.
All other variables are solved only for the zeroth order quantities.
Thus, the primary approximation in the low Mach number model is the
use of the base pressure in the equations of state and energy. Since
the equation of state involves a relationship between $\rho$ and
$T$, generally both the energy equation and the continuity equation
are not solved in the computational model. Rather one of these equations
is solved and the other employed as a constraint together with the
momentum equation to develop an elliptic equation for the pressure
perturbation $P^{\prime}$. Different choices have been described
in the literature (see below).

\begin{figure}
\plotone{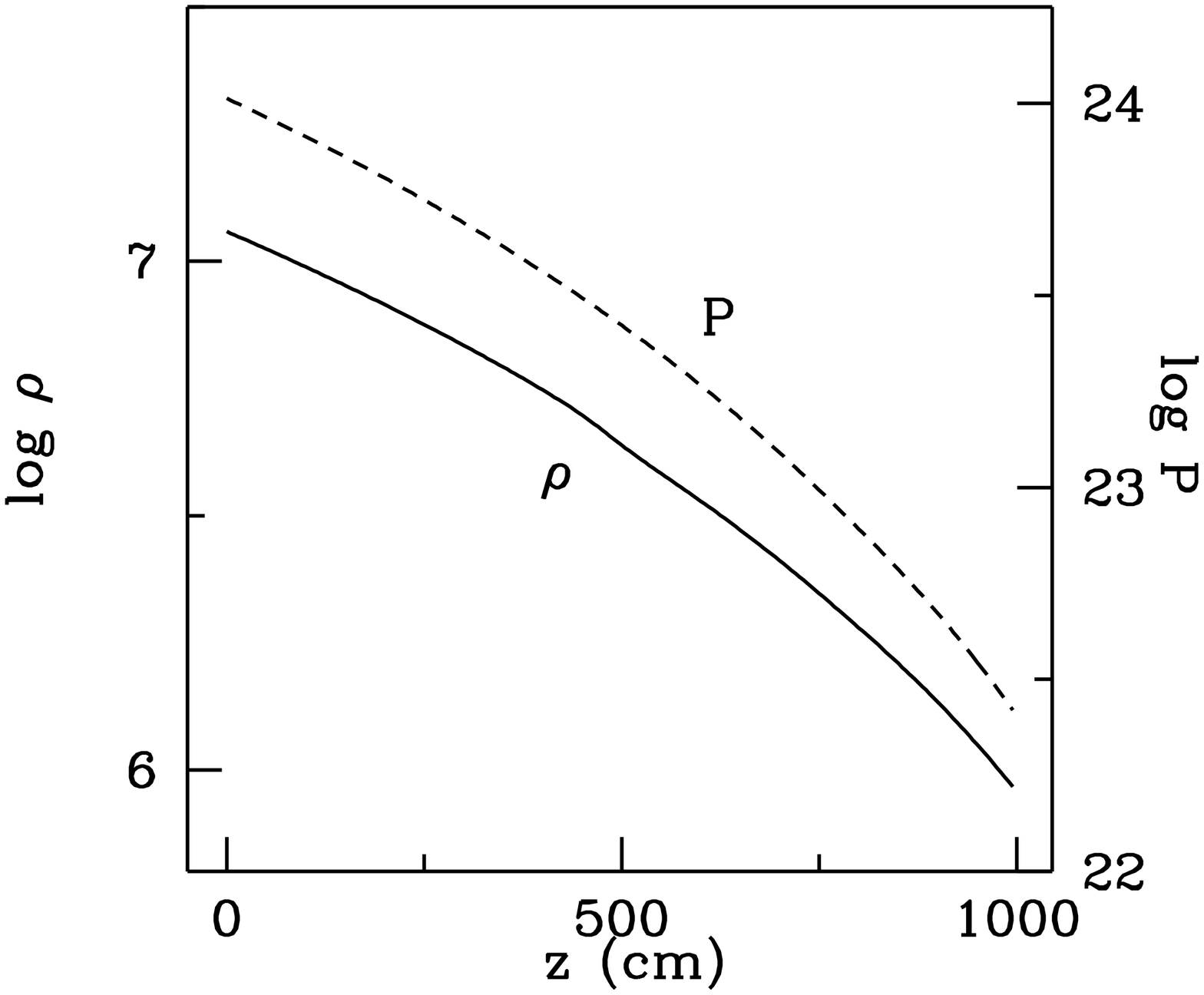}

\caption{{\small Initial $\rho$ and $P$. Initial vertical profile of the
2D domain. The solid line is $\rho$, measured on the left axis. The
dashed line is $P$, measured on the right axis. }}

\label{fig:init_rho_P} 
\end{figure}

In the neutron star configurations considered here, the deflagrations
do not occur under isobaric conditions due to large vertical pressure
stratifications, so a more appropriate base state is one of hydrostatic
equilibrium. In our model, all hydrodynamic quantities are expanded
around a precomputed hydrostatic base state, which is assumed to be
independent of time, although that assumption can be readily relaxed.
That is, for any of the hydrodynamic quantities,

\begin{equation}
A(\vec{r},t)=A_{h}(z)+A^{\prime}(\vec{r},t),\label{eq:pertrub}\end{equation}
 where $\vec{r}$ is the position vector, $t$ is time, $z$ is the
vertical direction, and $A_{h}(z)$ represents the hydrostatic base
state of the hydrodynamic quantity. Only the hydrostatic pressure
$P_{h}(z)$ is kept in the equation of state and the energy equation,
thereby filtering out sound waves. The equations of the model are
solved directly for the physical variables since due to the wide range
of scales involved (e.g., 22 $<$ $log\, P$ $<$ 24; ref. Fig. \ref{fig:init_rho_P}),
non-dimensionalization is not particularly useful. In the expansion
(eq. {[}\ref{eq:pertrub}]) the only perturbed quantity is $P^{\prime}$.
The fundamental assumption in the model is that $P^{\prime}$ is small
compared to $P_{h}$ and can be neglected in the equations of state
and energy. No assumptions are made regarding the magnitude of $\rho^{\prime}$
and $T^{\prime}$. They are defined as $\rho^{\prime}=\rho-\rho_{h}$
and $T^{\prime}=T-T_{h}$ respectively so that effectively we solve
for the full density and temperature in our model. The introduction
of the primed variables is simply a computational device to explicitly
incorporate the hydrostatic equilibrium base state and eliminate numerical
errors due to the cancellation of large quantities in the vertical
momentum equation. As an ex-post-facto justification of the assumption,
we verified that $M<0.15$ throughout the simulation and that $\mid P^{\prime}/P_{h}\mid\leq M^{2}$,
often decidedly less.

In principle, a laterally varying base state can be incorporated in
the model. However, in this case a lateral flow field would be set
up due to the lateral pressure gradients. Such a flow field would
then result in a vertical flow field so that the resulting base state
would no longer be in hydrostatic equilibrium. In view of the large
values of $g$, any significant imbalance between the pressure gradient
and buoyancy term in the vertical momentum equation would necessarily
result in large velocities inconsistent with the assumptions of the
low Mach number model. Hence, $P_{h}$ is laterally independent in
our LMNA model.

We next describe the equations of the model. By substituting equation
(\ref{eq:pertrub}) into the Euler equations, the LMNA governing equations
are obtained and presented in the order in which they are solved:

\begin{equation}
\frac{DX_{l}}{Dt}=R_{l},\label{eq:species}\end{equation}

\begin{equation}
\rho c_{p}\frac{DT}{Dt}-\delta\frac{DP_{h}}{Dt}=Q+\diver{\kappa\grad T}-\sum_{l}\rho\frac{\partial H}{\partial X_{l}}R_{l},\label{eq:energy}\end{equation}

\begin{equation}
\rho=F(T,P_{h},X_{l}),\label{eq:state}\end{equation}

\begin{equation}
\rho\frac{D\vel}{Dt}+\grad P^{\prime}=\rho^{\prime}\gvec.\label{eq:momentum}\end{equation}
 These expressions are the equations of species (eq. {[}\ref{eq:species}]),
energy (eq. {[}\ref{eq:energy}]), state (eq. {[}\ref{eq:state}]),
and momentum (eq. {[}\ref{eq:momentum}]). Here, $\frac{D}{Dt}$ is
the material derivative; $\gvec$ is the gravitational acceleration
vector ( = $-g\khat$ in the vertical direction), where $g$ is assumed
to be a constant, since its value changes by only 0.1\% over the vertical
extent presently modeled; $c_{p}$ is the specific heat at constant
pressure; $\delta$ is a thermodynamic coefficient = $-\left(\frac{\partial ln\rho}{\partial lnT}\right)_{P}$;
$Q$ is the $3\alpha$ rate of energy generation per unit volume;
$H$ is the enthalpy; $R_{l}$ is the rate of production and depletion
of species $l$; and $\kappa$ is the radiative thermal conductivity
= $\frac{4}{3}\frac{acT^{3}}{\kappa_{o}\rho}$, where $a$ is the
radiation constant, $c$ is the speed of light, and $\kappa_{o}$
is the radiative opacity, calculated from analytical expressions as
referenced in Iben (1975), Christy (1966), and Weaver et al. (1978).
The tabulated Helmholtz equation of state (Timmes \& Swesty 2000)
of an ionized gas includes the contributions of radiation as well
as electrons in an arbitrary state of degeneracy. The energy generation
rate (EGR) is obtained from the S-matrix calculations of Fushiki \&
Lamb (1987) which includes electron screening factors. Subsequent
$\alpha$-capture reactions involving $_{6}^{12}C$ are neglected,
since calculations show that the EGR of these reactions before the
peak of the burst are negligible compared to that of the $3\alpha$
reaction. Base state hydrostatic equilibrium ($\grad P_{h}=-\rho_{h}\gvec$
) is employed in equation (\ref{eq:energy}). Fluid viscosity is neglected
in view of the large Reynolds numbers expected on the neutron star.

On the right-hand-side of the energy equation (eq. {[}\ref{eq:energy}]),
$\sum_{l}\rho\frac{\partial H}{\partial X_{l}}R_{l}$ is the change
in energy due to the change in composition. We have investigated this
term and found that its effect is negligible. We therefore do not
include it in our computations due to the extra computational cost
required for its evaluation in the equation of state.

Note that $P^{\prime}$ appears only in the momentum equation (eq.
{[}\ref{eq:momentum}]). Generally, cancellation between the perturbed
pressure gradient and the $\rho^{\prime}\gvec$ buoyancy term occurs,
which can result in significant numerical errors, compounded by the
large values of $g$. The accuracy and stability of the computation
is significantly enhanced by introducing the variable

\begin{equation}
\phi=P^{\prime}-gK,\quad K=\int_{z_{top}}^{z}\rho^{\prime}dz^{\prime}\label{eq:phiK}\end{equation}
 where $K$ is an integrated density function, which is calculated
per lateral column from the top of the domain ($z_{top}$) downward.
Using $\phi$, the momentum equation (eq. {[}\ref{eq:momentum}])
can be equivalently expressed as

\begin{equation}
\rho\matder{\vel}=-\grad\phi-\dd{(gK)}{y}\jhat.\label{eq:momentumPHI}\end{equation}
 Note that $\grad\phi$ represents the net vertical force per unit
volume so that in this formulation, cancellation between the perturbed
pressure gradient and the $\rho^{\prime}\vec{g}$ buoyancy term is
explicitly accounted for. A term proportional to $g$ now enters into
the lateral momentum equation, which we have found does not cause
numerical difficulties.

By taking the divergence of the momentum equation (eq. {[}\ref{eq:momentum}])
and using the continuity equation ($\ddt{\rho}+\diver{\rho\vel}=0$),
an elliptic equation is obtained:

\begin{equation}
\grad^{2}\phi=\ddd{\rho}{t}-\diver\left\{ \diver(\rho\vel\vel)\right\} -\ddd{(gK)}{y}.\label{eq:elliptic}\end{equation}
 Solving the elliptic equation is necessary to close the set of governing
equations. Note that temperature is explicitly evolved with the energy
equation (eq. {[}\ref{eq:energy}]) but the continuity equation is
not solved directly. Density and temperature are related via the equation
of state for constant $P_{h}$. Since we enforce the equation of state,
density and temperature cannot be updated independently of one another.
As a result, either the continuity or energy equation can be included
in the equations of the model; the other equation is incorporated
in the compressibility constraint, that is, the use of the continuity
equation in the energy equation for terms involving the divergence
of the velocity field, ultimately leading to the right-hand side of
equation (\ref{eq:elliptic}). The equation of state is used to relate
density, temperature and mass fraction. (An alternative formulation,
relaxing the equation of state is described in Bell et al. 2004a.)
In terrestrial modeling, different approaches have been adopted in
the literature. Some approaches (e.g., McMurtry et. al. 1985; McGrattan
et. al. 1994, 2000, 2004, and the references therein) solve the continuity
equation and eliminate the energy equation, incorporating it into
the compressibility constraint. Other approaches (e.g., Majda and
Sethian 1985; Bayliss et al. 1992) solve the energy equation, employing
the continuity equation as a constraint. We adopt the latter approach.
The alternative formulation causes numerical difficulties in obtaining
the density from the equation of state due to the weak temperature
dependence of the pressure for the degenerate conditions that pertain
during the burst.

Using the continuity equation, the term $\ddd{\rho}{t}$ in equation
(\ref{eq:elliptic}) is numerically evaluated as:

\begin{equation}
\ddd{\rho}{t}=\frac{1}{\Delta t}\left[\left(\ddt{\rho}\right)^{n+1}+\diver(\rho\vel)^{n}\right],\label{eq:d2rhodt2}\end{equation}
 where $\Delta t$ is the time step, $n$ denotes values at the current
time-level, and $n+1$ denotes values at the next time-level. Here,
$\left(\ddt{\rho}\right)^{n+1}$ is determined from an analytical
expression for $\ddt{\rho}$ which can be obtained from evaluating
the partial derivatives of the internal energy. On the other hand,
$\diver(\rho\vel)^{n}$ is evaluated by centrally differencing local
values of $\rho$ and $\vel$ at the present time-level over two zones.
Solving equation (\ref{eq:d2rhodt2}) in this manner is required for
numerical stability.

The governing equations are solved with operator-splitting on a uniform,
staggered, plane-parallel, Cartesian grid which is parallelized using
the Message-Passing Interface (MPI) software. Upwind (advective and
first-order convective terms (terms having the form $\vel\cdot\grad\vel$,
see the Appendix below)) and central differencing (diffusion, $gK$,
and second-order convective terms) are used for spatial derivatives.
When needed, linear averaging is used, for instance, to find zone
center values from zone edge values. Forward-Euler is used to difference
time derivatives, using a Courant factor (the proportionality constant
which is required for numerical stability and relates the local timestep,
zone size, and propagation speed) of 0.50, which is comparable to
Courant factors employed in fully compressible 2D calculations. Thus,
the current model is first-order accurate in space and time, however,
a second order in time scheme can be easily implemented as a predictor-corrector
scheme (e.g., McGrattan et al 2004). In such a scheme a predicted
value is first computed at the new timestep. In the corrector step,
central differences in time are assumed analogous to the midpoint
method, thereby resulting in second order temporal accuracy. We have
tested such a predictor-corrector scheme and found that there was
only a negligible effect on the solution. Since the predictor-corrector
scheme essentially doubles the cost of the computation, we do not
use it in our computational scheme. This elliptic equation for $\phi$
(eq. {[}\ref{eq:elliptic}]) is solved using FISHPAK, a package of
subprograms for the solution of separable, elliptic partial differential
equations developed by Adams, Swarztrauber, and Sweet (1988). 

The timestep is determined adaptively from the solution. At each step,
the smallest timestep based on the flow speed ($dt_{CFL}$), thermal
diffusion, and nuclear burning in the entire domain is compared to
a fixed, maximum timestep $dt_{max}$ (a parameter, chosen to be 5$\times$10$^{-6}$
s), and the smallest one becomes the timestep for the next computational
step. Numerical stability requires using $dt_{max}$, and its necessity
is related to the large value of $g$ in this problem. Only during
the initial part of the simulation does $dt_{max}$ dominate over
the other timesteps, and for the remainder of the simulation, $dt_{CFL}$
dominates. Also, the maximum allowed percentage increase in timestep
is 10\% above the previous value.

For the initial conditions, 10$^{3}$ cm of a neutron star envelope
is constructed in thermal and hydrostatic equilibrium in the manner
of Bildsten (1995) with a 1 cm/zone resolution. Canonical neutron
star parameters are chosen: $M_{NS}=1.4M_{\odot}$, $R_{NS}=$1.0
$\times$ 10$^{6}$ cm, and $g$ = $\frac{GM_{NS}}{R_{NS}^{2}}$ =
2 $\times$ 10$^{14}$ cm s$^{-2}$. (We note that $g$ is large,
thereby warranting special numerical procedures as described above
to deal with cancellation between the pressure gradient and buoyancy
force in the vertical momentum equation of the 2D model.) The mass
accretion rate per unit area is $\dot{{m}}=\frac{5\times10^{-9}}{4\pi R_{NS}^{2}}M_{\odot}$
yr$^{-1}$cm$^{-2}$. We note that at such accretion rates, our assumption
of helium burning corresponds to the accretion of helium, rather than
a hydrogen rich composition. Such a circumstance could apply to the
case of ultra compact low mass X-ray binary systems (see, for example,
Nelemans \& Jonker 2006). The flux at the base of the model is chosen
to be 5 $\times$ 10$^{20}$ erg s$^{-1}$ cm$^{-2}$, a value corresponding
to crustal heating associated with residual nuclear burning. (Changing
this value affects the depth at which the burst occurs.) This initial
structure is used in the 2D model as the reference base state ($P_{h}$,
$T_{h}$, $\rho_{h}$). To generate the starting conditions for the
2D hydrodynamical evolution, the initial structure is first evolved
with a 1D diffusional-thermal code (which solves eqs. {[}\ref{eq:species}],
{[}\ref{eq:energy}], and {[}\ref{eq:state}]) through several burst
cycles with hydrostatic equilibrium re-calculated after each timestep.
From this, subadiabatic conditions which exist approximately 1 s prior
to the peak of the burst become the starting conditions for the 2D
model. This 1D structure is reproduced across the width of the computational
domain to create a laterally uniform 2D domain. While several domain
dimensions were tested, the domain used for the results presented
here has 386$\times$200 zones at a resolution of 5 cm/zone. To save
on computational expense, the upper boundary of the 2D domain is 500
cm below the actual surface of the star, which is not modeled.

\begin{figure}
\plotone{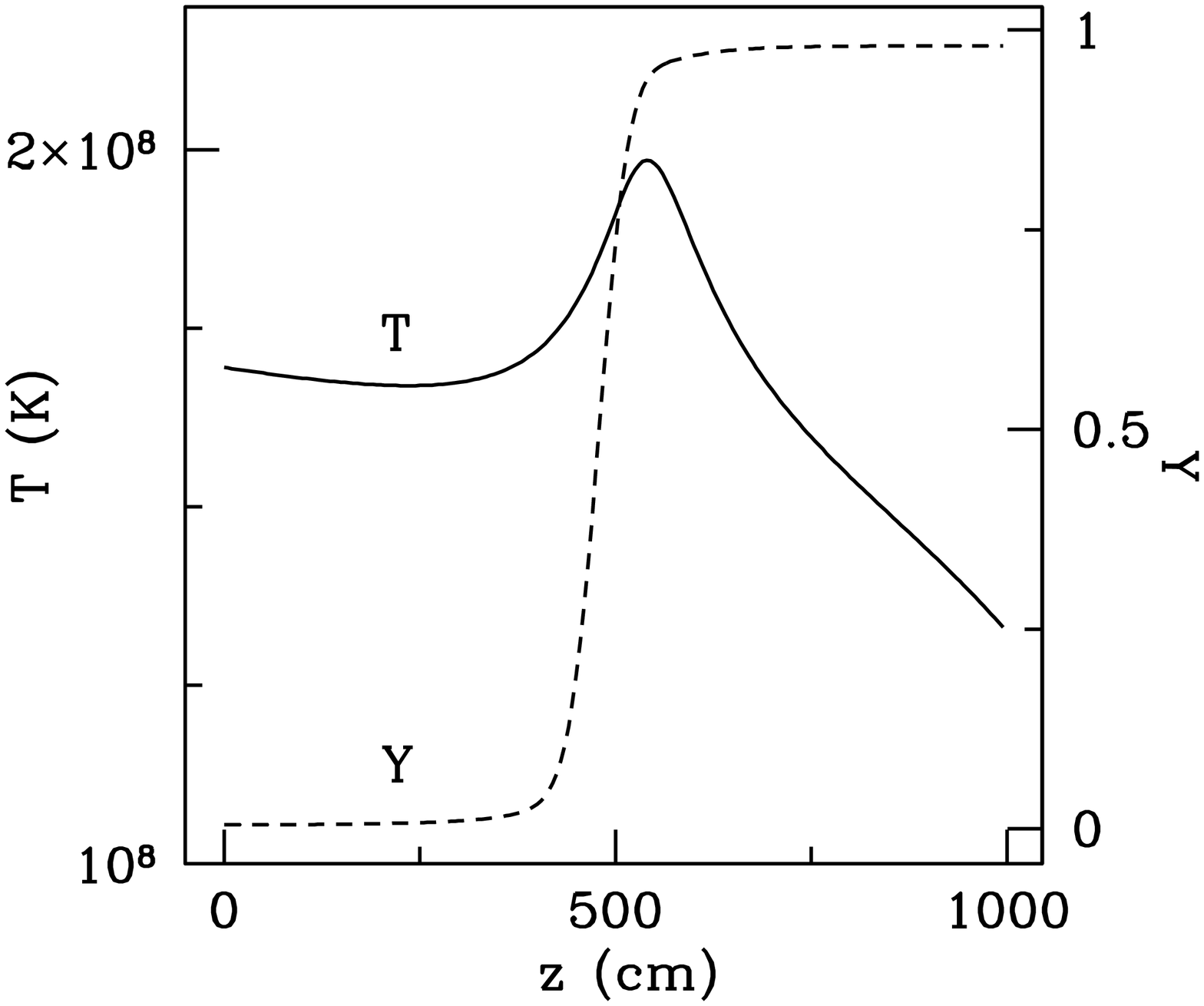}

\caption{{\small Initial $T$ and $Y$. Initial vertical profile of the 2D
domain. The solid line is $T$, measured on the left axis. The dashed
line is $Y$, measured on the right axis. }}

\label{fig:init_T_Y} 
\end{figure}

Initially, as shown in Figure \ref{fig:init_rho_P}, the domain extends
over one order of magnitude in $\rho$ (10$^{6}$- 10$^{7}$ g cm$^{-3}$)
and nearly two orders of magnitude in $P$ (10$^{22}$- 10$^{24}$
erg cm$^{-3}$) corresponding to 4.5 pressure scale heights ($H_{P}\sim200$
cm). Since thermal and compositional histories are taken into account
from the diffusional-thermal evolution, Figure \ref{fig:init_T_Y}
shows that the temperature profile has a local maximum and inversion
where the burst ignites from the hottest layer. The initial maximum
value of $T$ is 2$\times$10$^{8}$ K, which corresponds to an initial
maximum EGR of 7$\times$10$^{14}$ erg g$^{-1}$s$^{-1}$. Figure
\ref{fig:init_T_Y} also shows a very steep gradient in $Y$ ($_{2}^{4}He$
mass fraction) occurring where the hottest layer is located, such
that $Y\sim0.98$ above and $Y\sim0.02$ below. Because only two species
are considered, the ash complement is $Z$ ($_{6}^{12}C$ mass fraction).

It is important to note that the hottest burning layer corresponds
to a density of $\sim$ 4 $\times$ 10$^{6}$ g cm$^{-3}$, and that
the density at the top of the computational domain is of order 10$^{6}$
g cm$^{-3}$. Thus, we do not model the lower density regions of the
neutron star's upper atmosphere, a region which extends 500 cm above
the top of our computational domain. This region is not modeled for
several reasons: 1) it requires a significantly smaller timestep,
due to the timestep restriction associated with thermal diffusion,
which reduces the time-savings advantage afforded by the LMNA approximation;
2) its lower densities facilitate the development of larger magnitude
flows and hence larger Mach numbers, thus limiting the applicability
of the LMNA model there; 3) its degeneracy is lifted at advanced stages
of the burst, resulting in significant vertical expansion (e.g. Cumming
and Bildsten 2000), thus requiring a time-dependent pressure base
state to properly model, a feature which is not incorporated into
the current LMNA model, but which can be readily developed into future
models. Because the current model examines relatively high density
regions at depth, no significant expansion occurs in the model throughout
most of the burst event, and a time-independent base state is sufficient.

To provide lateral inhomogeneity, a small, 2D Gaussian perturbation
in density is applied ($\left(\frac{\Delta\rho}{\rho}\right)_{max}$=$-$1$\times$10$^{-6}$,
full-width-half-max = 50 cm), such that the center of the perturbation
is laterally centered and vertically positioned at the latitude of
the hottest layer. Without this perturbation, the burst proceeds in
a laterally homogeneous manner, and no 2D convective structures form.
Thus, the purpose of the initial perturbation is merely to break the
lateral symmetry of the starting conditions, allowing 2D dynamical
structures to naturally develop during the subsequent evolution. As
long as the magnitude of the perturbation in $\rho$ is less than
a fractional difference of 10$^{-2}$, the subsequent dynamics is
qualitatively independent of the magnitude and placement of the initial
perturbation in the domain.

Periodicity is assumed at the lateral domain edges for all evolved
variables. As verified by extensive numerical experiments, the domain's
lateral size is sufficiently wide so as not to affect the size and
evolution of the convection cells which naturally develop. At the
lower boundary, $\ddz{T}$, $\ddz{\rho},$$\ddz{X_{l}}$, and $\ddz{v}$
are all set to zero, where $v$ is the lateral component of the velocity.
The vertical component of velocity $w$ and all convective terms are
set to zero. In the elliptic solver, the Neumann condition ($\frac{\partial\phi}{\partial z}$=
$0$) is used, which is equivalent to imposing hydrostatic equilibrium
at the lower boundary. The domain's lower edge is set sufficiently
deep such that flows which develop there are relatively small ($|\mathbf{v}|<10^{2}$
cm s$^{-1}$) compared to those in the convective layer ($10^{4}<|\mathbf{v}|<10^{7}$
cm s$^{-1}$). At the upper boundary, $\ddz{v}$, $\ddz{w}$, and
$\ddz{X_{l}}$ are set to zero, and all convective terms are set to
zero. A temperature flux condition $\frac{\partial T}{\partial z}=-\frac{\rho T}{4N_{top}}$
is used to update $T$, where $N_{top}$ is the column density at
the upper boundary ($N_{top}=\frac{P_{top}}{g}$). The temperature
flux condition is derived from equating the radiative flux $F=-\frac{4ac}{3\kappa\rho}T^{3}\frac{dT}{dz}$
with the surface flux $F=\sigma T_{eff}^{4}$, where $T^{4}=\frac{3}{4}T_{eff}^{4}(\tau+\frac{2}{3})$
and $\tau=\int\kappa\rho dz=\frac{\kappa P}{g}$. Here, $T_{eff}$
is the effective temperature at the neutron star's surface. Also,
$F$ and $\kappa$ are assumed to be constant in the region between
the upper domain boundary and the actual surface of the star and that
this region is in radiative equilibrium. A consistent $\rho$ is then
found via the equation of state. In the elliptic solver, the Dirichlet
condition ($\phi$= $0$) is imposed. Since $K=0$ at the upper boundary
(see eq. {[}\ref{eq:phiK}]), the Dirichlet condition is effectively
a condition that $P^{'}=0$ at the top. We note that the results are
insensitive to the placement of the lower boundary, provided that
it is sufficiently deep. Likewise, the results are insensitive to
the location of the upper boundary, provided it is sufficiently high,
except for a very weak sensitivity near burst peak. The boundary conditions
described above allow for the burst and do not result in any noticeable
numerical oscillations or numerical instabilities.

A more detailed description of the computational model is given in
the Appendix, as well as in Lin (2006).

\section{Type I X-ray burst deflagrations }

\subsection{Convective dynamics}

The important thermodynamic and hydrodynamic features which develop
during the burst calculation are first described. By way of definition,
the subscript \textit{max} indicates the instantaneous spatial maximum
value of a quantity; the subscript \textit{peak} indicates the greatest
value of the quantity throughout the entire burst sequence; and the
subscript \textit{ave} indicates the spatially-averaged value of the
quantity over a particular analysis period.

\begin{figure}
\plotone{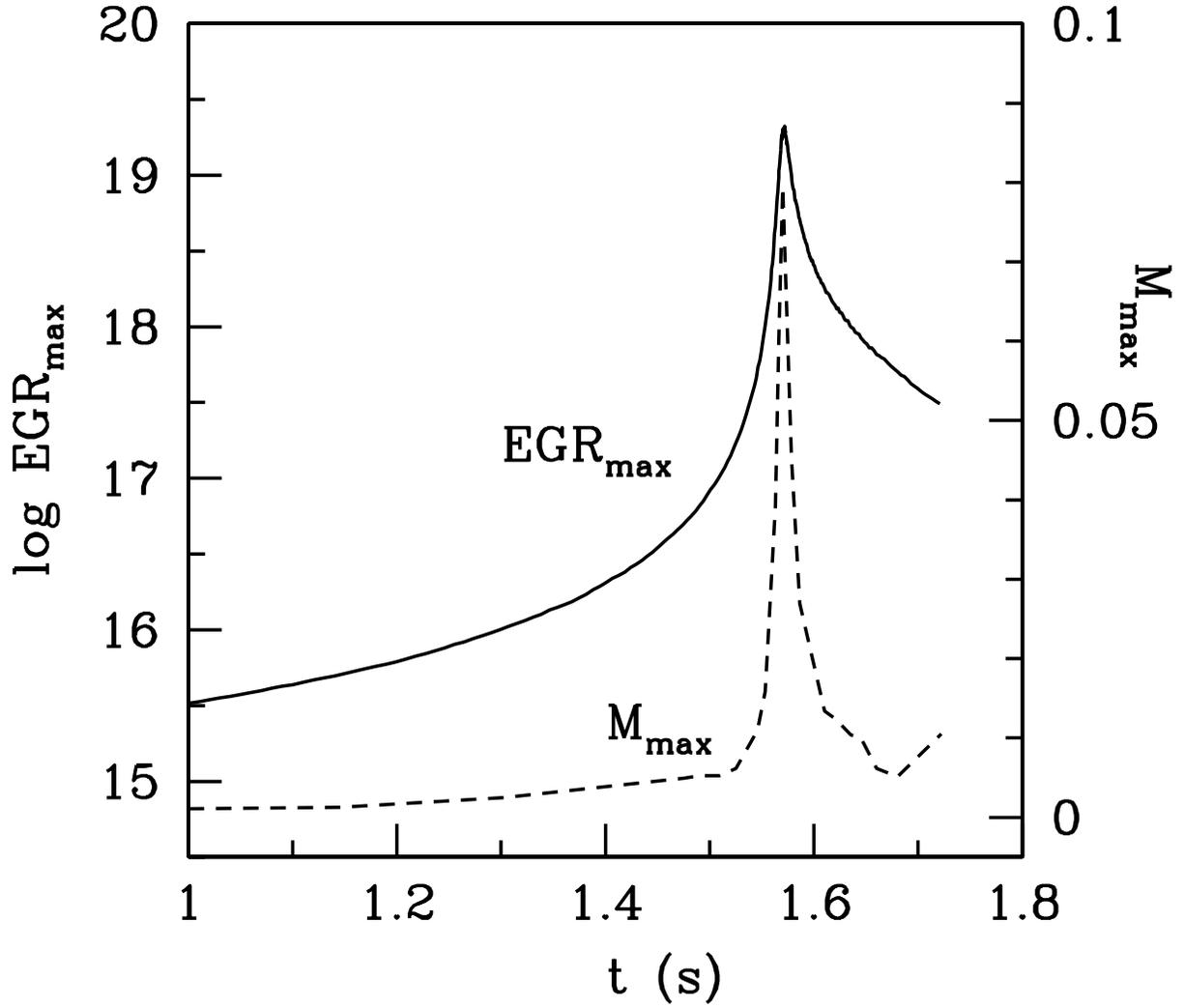}

\caption{{\small $log$  EGR$_{max}$ and $M_{max}$ vs. time (s). Instantaneous
maximum values of EGR (solid line, left axis) and Mach number $M$
(dashed line, right axis) over the entire domain as a function of
physical time from the start of the calculation ($t$ = 0 s). Note
the peak value of the Mach number is always less than 0.15.}}

\label{fig:egr_mach_time} 
\end{figure}

\begin{figure}
\epsscale{0.90} \plotone{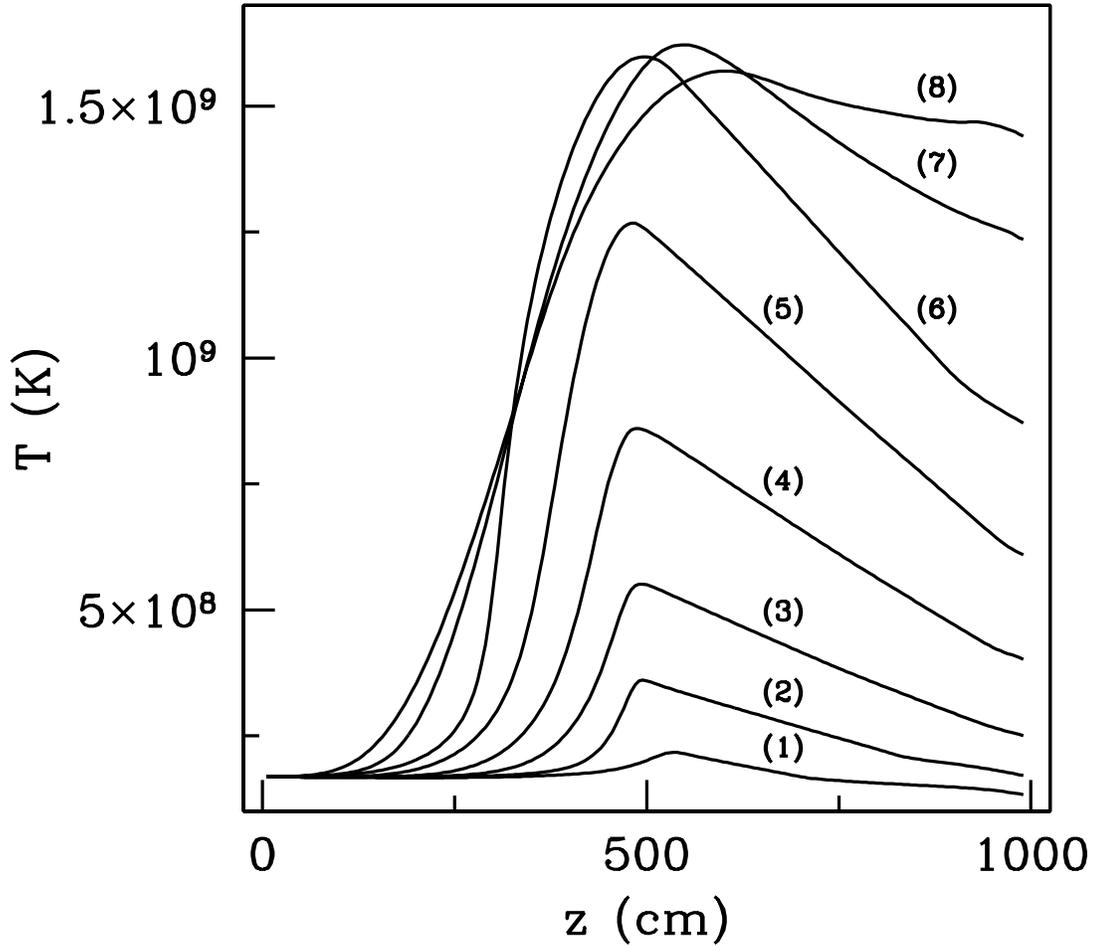}

\caption{{\small \label{fig:Tprofile}Temperature profile at several times.
Pre-burst peak, (1) corresponds to $log$  EGR$_{max}$ = 15.5, (2)
18.0, and (3) 19.0. At burst peak, (4) corresponds to $log$  EGR$_{peak}$
 = 19.3. Post-burst peak, (5) corresponds to $log$  EGR$_{max}$
= 18.8, (6) 18.0, (7) 17.5, and (8) 17.4.}}
\end{figure}

The value of EGR$_{max}$ quantifies the burst evolution sequence,
as shown in Figure \ref{fig:egr_mach_time}, where EGR$_{max}$ (solid
line) is plotted as a function of time. The time $t$ is given in
terms of the physical time after the start of the calculation ( $t$
= 0 s). Throughout the burst sequence, EGR$_{max}$ can be found in
a layer, henceforth referred to as the \textit{burning layer}, the
height of which is roughly constant throughout the burst progression
at approximately 500 cm above the domain base and which corresponds
to the height of the initially hottest layer of the starting conditions.
At the peak of the burst, $t$ = 1.572 s and the peak value of the
EGR during the entire sequence is EGR$_{peak}$ = 2 $\times$ 10$^{19}$
erg g$^{-1}$ s$^{-1}$. Simultaneously, the maximum flow speed also
peaks at 5$\times$10$^{7}$ cm s$^{-1}$ ($M_{peak}$ = 0.15; Fig.\ref{fig:egr_mach_time}
dashed line), which justifies using the LMNA to model this phenomenon.
Fig. \ref{fig:Tprofile} shows the laterally averaged temperature
as a function of height. As will be quantified below, the temperature
gradient is nearly adiabatic in the vertically expanding layer where
convection develops, and after the burst peak, adiabaticity is lost
since the temperature profile eventually becomes more uniform in the
upper domain resulting from the upward transport of heat by thermal
diffusion. Due to heating terms still dominating cooling terms at
the time of burst peak, the temperature peaks $\sim$ 0.10 s after
the EGR peaks ($T_{peak}$ = 1.7$\times$10$^{9}$ K).

Initially, everything is quiescent, since $\grad<\grad_{ad}$ everywhere,
where $\grad\equiv\left(\frac{d\, ln\, T}{d\, ln\, P}\right)_{actual}$
and $\grad_{ad}\equiv\left(\frac{d\, ln\, T}{d\, ln\, P}\right)_{s}$,
which is the adiabatic temperature gradient (Schwarzschild 1906).
Lateral inhomogeneities introduced by the initial perturbation cause
small-scale, localized eddies to develop everywhere in the domain.
The magnitude of $\vel_{ave}$ of these eddies is on the order of
10$^{2}$ cm s$^{-1}$. At $t$ = 0.15 s ($log$  EGR$_{max}$ = 14.9),
a thin layer approximately 50 cm in vertical extent becomes superadiabatic,
such that $\left(\Delta\nabla\right)_{max}$ = 0.12 (where $\Delta\nabla\equiv\grad-\grad_{ad}$,
the adiabatic excess), and by $t$ = 0.35 s ($log$  EGR$_{max}$
= 15), convective motions naturally develop and become evident in
this layer. At this time, the magnitude of $\vel_{ave}$ in the convective
region is of order 10$^{4}$ cm s$^{-1}$, however the flow field
initially lacks apparent structure. As the burst proceeds and EGR$_{max}$
rises, the boundaries of the convective layer expand vertically due
to thermal diffusion of heat away from the burning layer. By $t$
= 0.70 s ($log$  EGR$_{max}$ = 15.2), the vertical extent of the
convective layer has increased to $\sim$ 150 cm, and the flow field
self-organizes into distinct Bénard cells (Bénard 1900; Koschmieder
1993), each characterized by central upflows and adjacent downflows.
The shapes and sizes of the convective cells are dynamically evolving
on subconvective timescales, which varies from 1000 $\mu s$ (at $log$
 EGR$_{max}$ = 16) to 150 $\mu s$ (at $log$  EGR$_{max}$ = 19).

\begin{figure}
\epsscale{0.78} \plotone{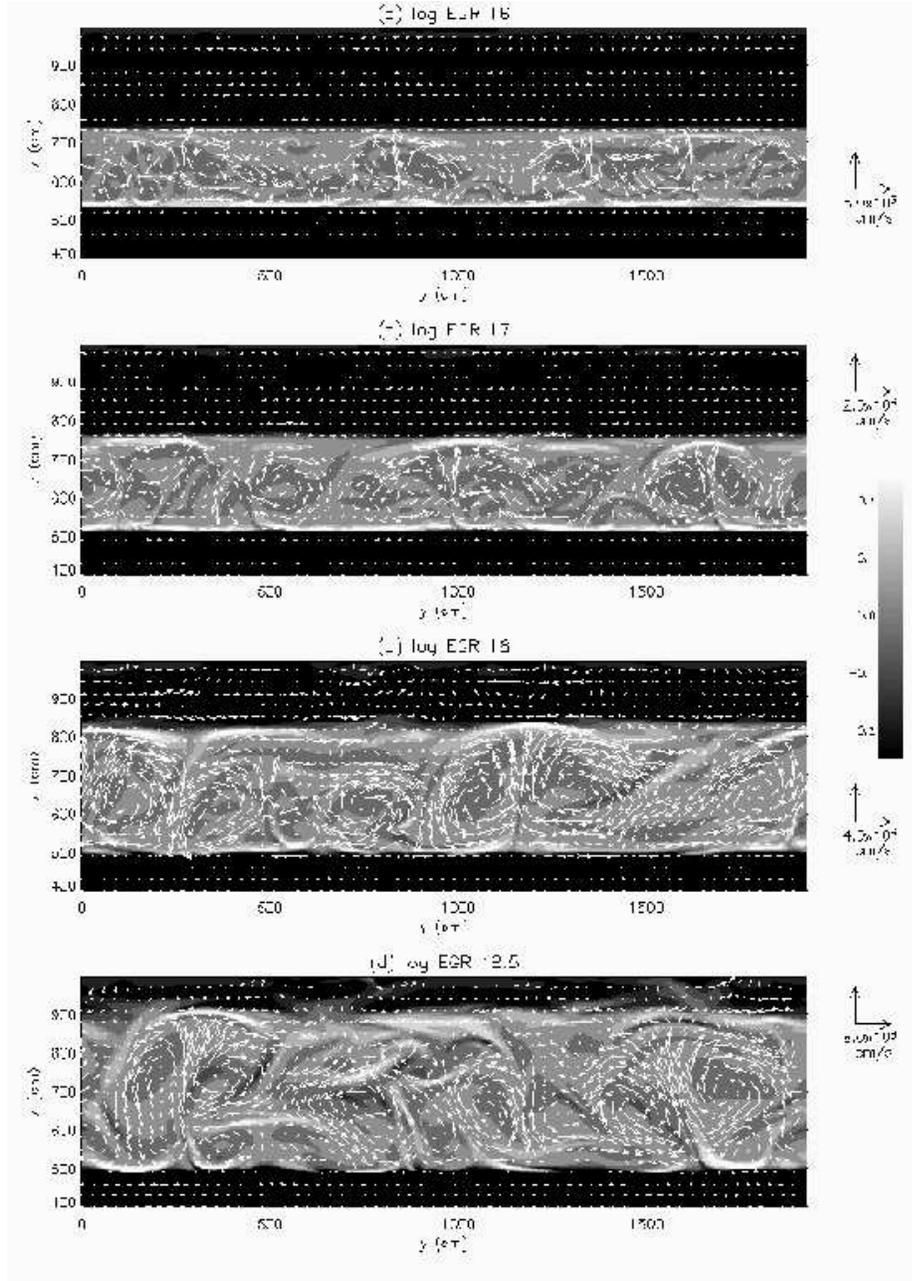}

\caption{{\small Flow fields at four stages of the burst, corresponding to
$log$  EGR$_{max}$ = 16 (a), 17 (b), 18 (c), and 18.5 (d). Velocity
vectors are superimposed over contours of $\Delta\nabla$}, {\small the
adiabatic excess. Vertical and lateral coordinates given with respect
to the lower left corner ($y=0,$ $z=0$ ) of the domain. For clarity,
every 6th velocity vector is plotted. }}

\label{fig:flowfield} 
\end{figure}

Figure \ref{fig:flowfield} shows flow fields superimposed against
contours of superadiabaticity ($\Delta\nabla$) in the domain at four
stages prior to burst peak. For each plot in the figure, the vertical
axis represents the vertical direction $z$ (cm), and the horizontal
axis measures the lateral direction $y$ (cm). (The origin of the
coordinate system is located at the lower left corner of the computational
domain. The entire lateral width of the domain is shown in the figure,
while the relatively calm, lower half of the domain is omitted for
clarity.) To enhance the clarity of the main features in the flow
fields, the resolution of the vector field is 30 cm/zone, that is,
every 6th velocity vector is plotted.

At $log$  EGR$_{max}$  = 16 and 17 (Figs. \ref{fig:flowfield}$a$
and \ref{fig:flowfield}$b$), distinct Bénard-like cells have self-organized,
and the larger cells are approximately symmetric with uniform aspect
ratios. The larger cells are characterized by central updrafts, which
channel material from the base of the convective layer through vertical
chimneys upward toward the top of the convective layer boundary. There,
the flows laterally bifurcate to form more diffuse downdrafts. The
flow fields dynamically evolve on a subconvective timescale. For example,
a given larger cell may break up into smaller cells. Smaller cells
may merge to form a larger cell. The direction of updrafts and downdrafts
may also skew diagonally. The aspect ratio of a cell may momentarily
change significantly, as the cell takes on a variety of shapes and
forms, but maintaining its overall outline as a complete cell. An
updraft occurring at one time at a certain lateral location may develop
into a downdraft at the same location within a few convective times.
No bulk lateral motion of the cells is apparent, as the individual
cells do not hold their integrity long enough to noticeably drift
as a unit. By the later stages of the burst ($log$  EGR$_{max}$
= 18 and 18.5; Fig. \ref{fig:flowfield}$c$ and \ref{fig:flowfield}$d$),
the convective layer noticeably expands due to thermal diffusion of
heat away from the burning layer. The boundary of the upper convective
layer propagates upward on the order of 10$^{4}$ cm s$^{-1}$, while
the lower boundary moves downward at 10$^{2}$ cm s$^{-1}$. As the
layer expands, the heights of the larger cells grow to fill up the
vertical extent of the layer. Correspondingly, the horizontal widths
of the larger cells also increase to maintain a roughly uniform aspect
ratio. The cells continue to dynamically morph into a variety of shapes,
orientations, and sizes on subconvective timescales.

Standing out in bold contrast against the radiative regions above
and below it, the convective layer can be seen in Figures \ref{fig:flowfield}
$a-d$ to be superadiabatic on average, but with subadiabatic pockets
within it. These subadiabatic regions can always be found in the convective
layer throughout the evolution and become more distinct at higher
EGR$_{max}$ levels. The most superadiabatic regions appear to form
near the bottom of the convective layer as elongated fronts which
are carried up by the strongest currents to the top of the convective
region. Laterally averaged over time, however, the convective layer
is slightly superadiabatic (+0.01 pre-burst peak, +0.001 post-burst
peak).

\begin{figure}
\plotone{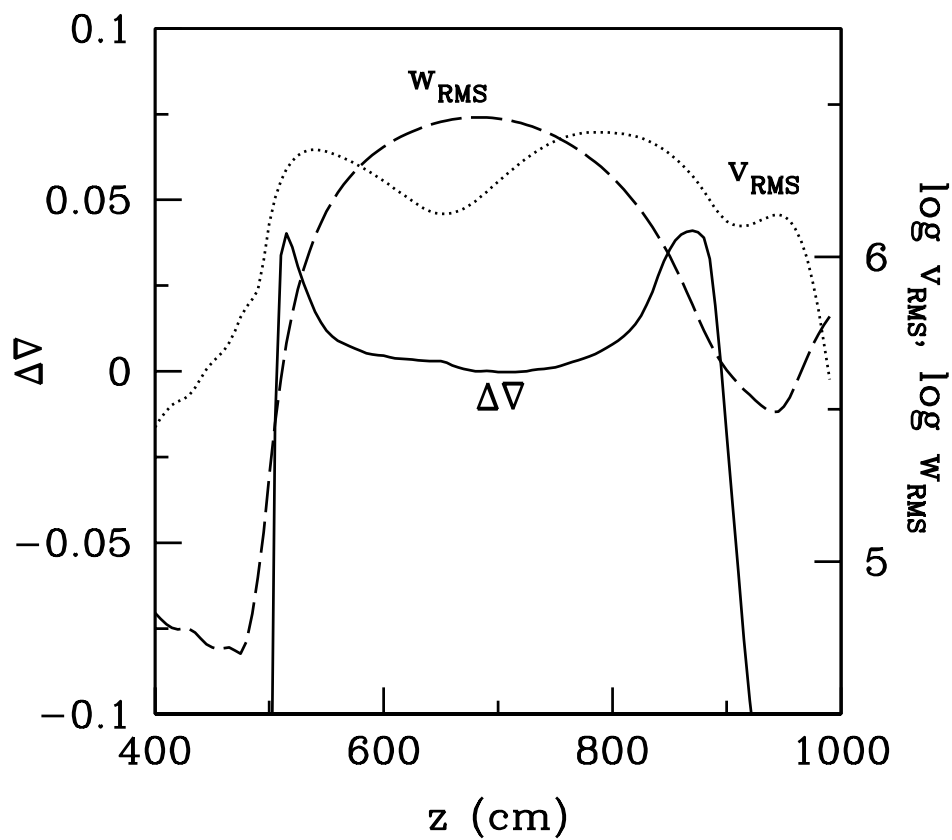}

\caption{{\small Adiabatic excess ($\Delta\grad$), $log$  $v_{RMS}$, and
$log$  $w_{RMS}$ vs. z (cm) at $log$  EGR$_{max}$ = 18.5. The
left axis measures $\Delta\grad$ (solid line); the right axis measures
$log$  $v_{RMS}$ (dotted line) and $log$  $w_{RMS}$ (dashed line).
The magnitudes of vertical speeds range by an order of magnitude within
the convective layer ($\Delta\grad>0$). At the convective layer boundaries,
lateral speeds exceed vertical speeds by an order of magnitude.}}

\label{fig:vw_profile} 
\end{figure}

Figure \ref{fig:vw_profile} illustrates the laterally and temporally
averaged, root-mean-square (RMS) values of $v$ and $w$ as functions
of height at $log$  EGR$_{max}$  = 18.5. For reference, the adiabatic
excess is also plotted. At this particular stage in the burst evolution
within the convective layer ($\Delta\grad>0$), $w_{RMS}$ peaks near
the center of the layer, and diminishes by an order of magnitude at
the layer boundaries. On the other hand, a local minimum in $v_{RMS}$
occurs near the center of the layer, while $v_{RMS}$ peaks toward
the layer boundaries. This is consistent with the general nature of
the flow in a Bénard-like cell. Notably, Figure \ref{fig:vw_profile}
shows that on average, $v_{RMS}$ exceeds $w_{RMS}$ at the boundaries
by nearly an order of magnitude. At other stages of the burst, the
qualitative behavior of these quantities as a function of height is
identical, but the range in magnitude may be as great as two orders
of magnitude (e.g., when $log$  EGR$_{max}$ $<$ 17). The dominance
of lateral over vertical flows at the layer boundaries helps to explain
the limited amount of tracer particle penetration beyond the convective
layer boundaries (see below).

Figure \ref{fig:benard_cell_17} focuses on a Bénard-like cell at
$log$  EGR$_{max}$ = 17. Here, the flow field is superimposed against
contours of temperature fluctuation, expressed with the notation\[
\Delta T\equiv T-T_{ave}(z),\]
 where $T$ is the instantaneous temperature at a given zone, and
$T_{ave}(z)$ is the lateral average at height $z$. (The fluctuation
in $_{2}^{4}He$ mass fraction $\Delta Y$ is similarly defined.)
Black velocity vectors are used where $w>0$, while white vectors
indicate where $w<0$. The cell in Figure \ref{fig:benard_cell_17}
exemplifies how upflows are generally associated with columns which
are relatively warmer than their surroundings. Conversely, downflows
are generally associated with slightly cooler regions. Regions which
are relatively warmer are necessarily lower in density, and therefore
rise due to buoyancy, resulting in upward flows, and vice versa. Upflows
are generally more collimated than downflows. The flow fields and
fluctuations vary on the same subconvective timescales. Moreover,
$\Delta T$ and $\Delta Y$ are complementary at all times, such that
up- and downflows are consistently characterized by deficits and excesses
in $Y$ respectively. As the flow speeds increase, so do the magnitude
of the fluctuations, and the maximum relative magnitude of these fluctuations
throughout the entire burst is $\pm0.10$.

\begin{figure}
\plotone{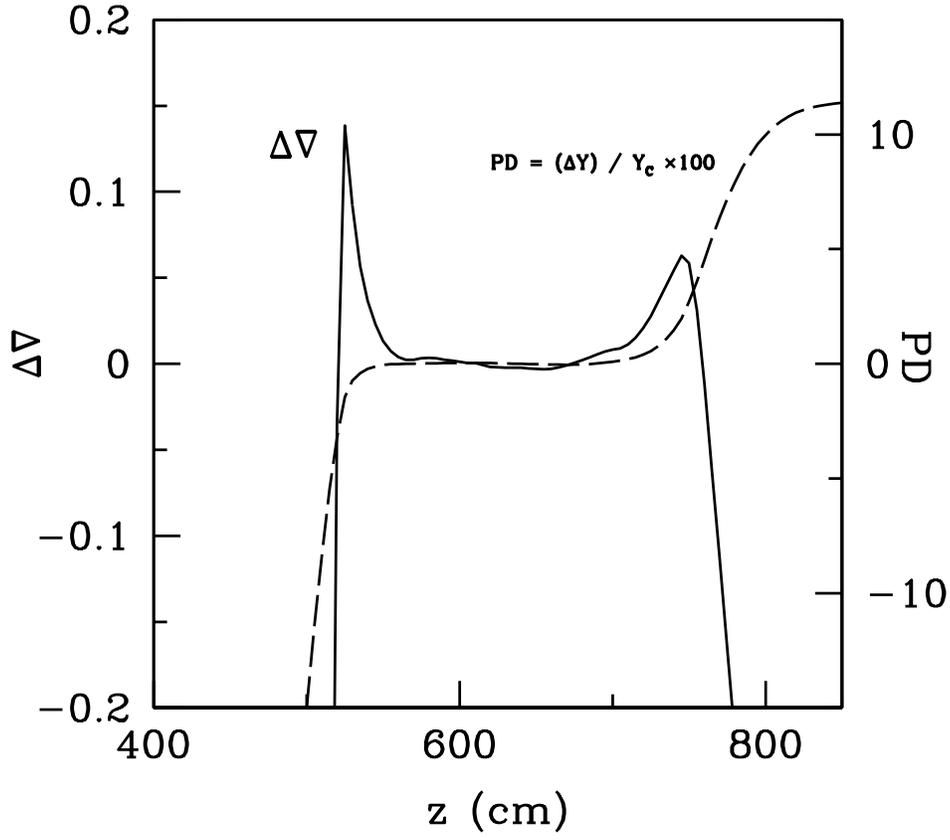}

\caption{{\small Profiles of percentage difference (PD) in $Y$ and adiabatic
excess ($\Delta\grad$) at $log$  EGR$_{max}$  = 17. The left axis
and solid line describe $\Delta\grad$, while the right axis and dashed
line describe PD. PD is taken with respect to $Y_{C}$, corresponding
to $Y$ at z = 625 cm, the center of the convective layer. While most
of the convective layer ($\Delta\grad>0$) is well-mixed, significant
composition gradients exist at the layer boundaries due to less efficient
mixing.}}

\label{fig:Y17profile} 
\end{figure}

The correlation between the sign of $\Delta Y$ and the flow direction
can be understood to be entirely an advective effect, since the timescales
over which these fluctuations occur ( $\sim$ 10$^{-5}$ - 10$^{-3}$
s) are many orders of magnitude smaller than burning timescales (
$\sim$ 10$^{-1}$ s). To help explain this association, Figure \ref{fig:Y17profile}
illustrates the profile of the percentage difference (PD = $\Delta Y/Y_{C}$,
where $\Delta Y=Y-Y_{C}$) of the laterally averaged value of $Y$
with respect to $Y_{C}$, the value at the center of the convective
region ($z$ = 625 cm) at $log$  EGR$_{max}$  = 17. As the figure
shows, convective mixing very efficiently homogenizes the composition
in most of the convective layer, where $Y$ varies by less than one
percent over a height corresponding to one pressure scale height at
this particular stage in the burst, and up to two pressure scale heights
at more advanced stages. However, at the layer's upper and lower boundaries,
substantial composition gradients exist due to less efficient mixing
there ( $\sim$ $-$5\% at the lower boundary, +5\% at the upper boundary).
Lateral flows also dominate in these regions, as previously described.
Thus, converging lateral flows at the sources of downflows locally
concentrate fuel within a relatively rich helium layer near the convective
layer's upper boundary, and the vertical flows advect fuel downward.
In the same manner at the lower boundary, local concentration of carbon
($Y$ deficit) occurs at the source of upflows, where carbon is advected
upward.

\begin{figure}
\plotone{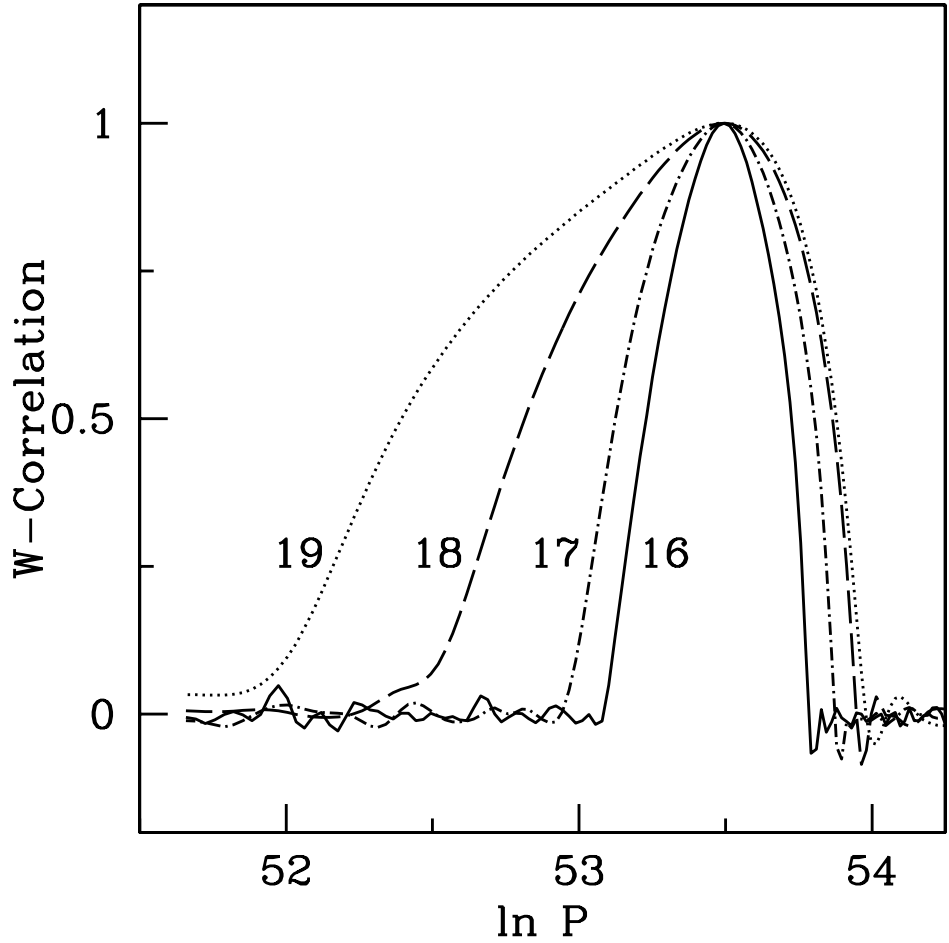}

\caption{{\small $W$ vs. $ln\, P$. Taken at four stages during the burst
progression: $log$  EGR$_{max}$ = 16 (solid line), 17 (dash-dot-dashed
line), 18 (dashed line), and 19 (dotted line). The height of the convective
layer increases to two pressure scale heights by $log$  EGR$_{max}$
= 19. The surface of the star is toward the left of the figure (lower
pressures).}}

\label{fig:W_progression} 
\end{figure}

To help quantify the extent and evolution of the convective layer,
the vertical velocity correlation function $W$ is calculated in the
manner of Chan \& Sofia (1987):

\[
W=\frac{\left\langle w_{k}w_{ref}\right\rangle }{\langle w_{k}\rangle^{1/2}\langle w_{ref}\rangle^{1/2}}.\]
 The vertical velocity correlation $W$ is constructed by laterally
and temporally averaging the product of $w$ at two vertical positions,
one of which is a fixed reference position ($w_{ref}$). For each
EGR level, this reference position is taken to be at the same vertical
height, corresponding to the center of the convective layer at $log$
 EGR$_{max}$ = 16. $W$ has a bell-shaped structure, and it can be
used to quantitatively define the positions of the vertical edges
of the larger Bénard cells. With reference to both the Schwarzschild
criteria ($\grad>\grad_{ad}$) and the actual sizes of the larger
cells which develop, the convective layer can be characterized by
regions where $W$ $>$ 0.10. For example, Figure \ref{fig:W_progression}
shows the vertical expansion of the convective layer as represented
by $W$ plotted as a function of $ln\, P$ at four EGR levels leading
up to the peak of the burst. By $t$ = 1.5 s ($log$  EGR$_{max}$
= 17), the vertical extent of the convective region is roughly one
pressure scale height, and by burst peak at $t$ = 1.572 s ($log$
 EGR$_{max}$ = 19.2), it has expanded to about two pressure scale
heights. Also, the lateral velocity correlation function $V$ is calculated
using a reference position at the lateral center of the domain. Comparing
$V$ with $W$ shows that the ratio of lateral to vertical extents
of the larger cells throughout the burst evolution varies between
0.9 and 1.1, such that on average, they feature a roughly uniform
aspect ratio.

\begin{figure}
\plotone{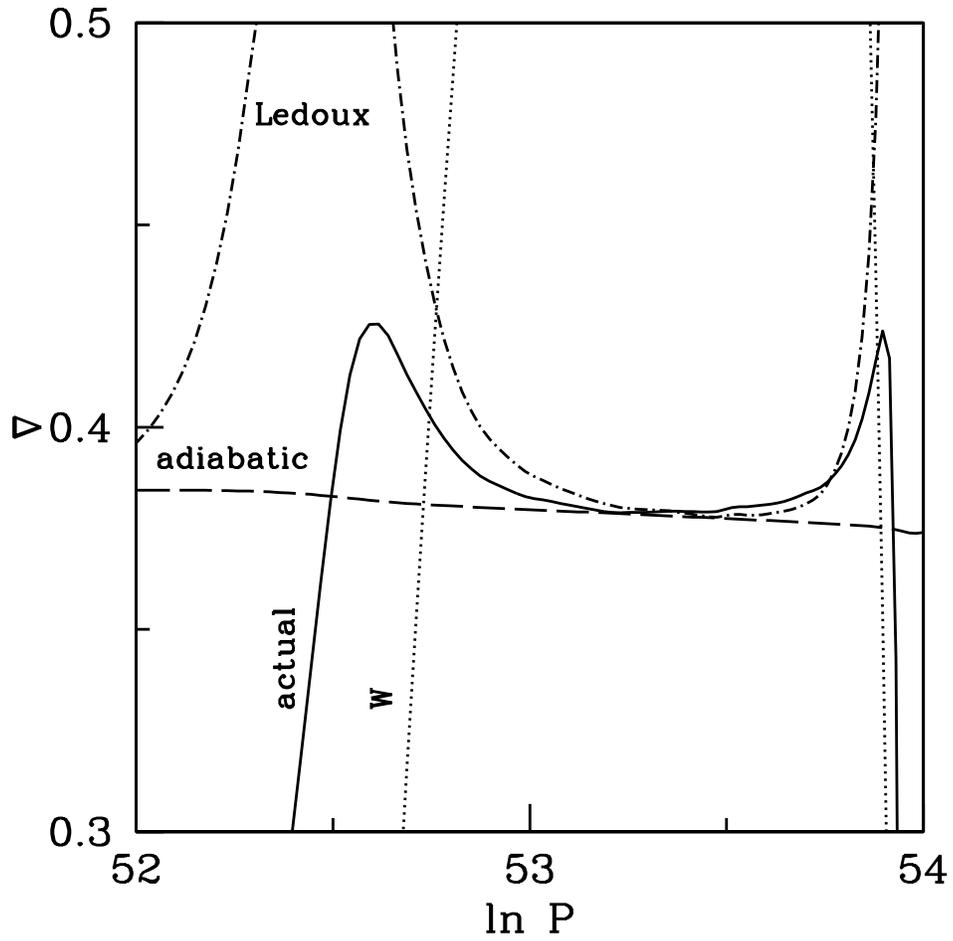}

\caption{{\small Thermodynamic gradients and $W$ vs. $ln\, P$ at $log$
 EGR$_{max}$ = 18. The solid line is $\grad$, dashed line is $\grad_{ad}$,
dash-dot-dashed line is $\grad_{L}$, and dotted line is $W$. The
convective region is characterized by a slight adiabatic excess. Near
the convective layer boundaries, $\grad$ more closely follows $\grad_{L}$.
The surface of the star is toward the left of the figure (lower pressures).}}

\label{fig:W_18} 
\end{figure}

Figure \ref{fig:W_18} shows $W$, $\grad$, $\grad_{ad}$, and the
Ledoux gradient $\grad_{L}\equiv\grad_{ad}+\frac{c_{1}}{c_{2}}\,\grad_{\mu}$
as a function of $ln\, P$ at $log$  EGR$_{max}$ = 18, where $\grad_{\mu}\equiv\left(\frac{d\, ln\,\mu}{d\, ln\, P}\right)$,
$c_{1}\equiv\left(\frac{\partial\, ln\,\rho}{\partial\, ln\,\mu}\right)_{P,T}$,
$c_{2}\equiv-\left(\frac{\partial\, ln\,\rho}{\partial\, ln\, T}\right)_{P,\mu}$,
and $\mu$ is the mean molecular weight. Here, as is true at all EGR
levels, the convective layer is superadiabatic on average. Moreover,
the actual temperature gradient more closely follows the Ledoux gradient
near the convective layer boundaries, due to steep composition gradients
which persist because convective mixing is less efficient there. The
figure also shows a region near the base of the convective layer which
satisfies the semiconvective criteria, $\grad>\grad_{L}$ (Ledoux
1947). This region consistently satisfies the semiconvective criteria
when $log$  EGR$_{max}$ $>$ 18.

Post-burst peak, the superadiabaticity of the convective layer decreases
as more of the domain heats up and the temperature gradient moderates.
Consequently, convective motions gradually diminish, and the Mach
number falls. By the conclusion of the calculation ($t$ = 1.74 s),
the domain has become completely subadiabatic again, and all convective
motions cease. Residual motion occurs at the upper boundary, but these
velocities are relatively insignificant ($M_{max}$ $<$ 0.02).

\subsection{Convective material transport}

As the burst evolves, mixing of composition from radiative regions
into the convective layer naturally occurs on the timescale of the
burst ($\sim$ 0.1 s) due to the vertical expansion of the convective
layer caused by thermal diffusion of heat away from the burning layer.
Compared to the burst timescale, the convective timescale is many
orders of magnitude smaller ($\sim$ 1000 $\mu s$ at $log$  EGR$_{max}$
= 16; $\sim$ 150 $\mu s$ at $log$  EGR$_{max}$ = 19). Thus, a
question which arises is whether significant penetration through the
formal boundaries of the convective layer occurs on convective timescales.
To address this question, transport of test particles through the
convective layer boundaries is studied from two perspectives, transport
out of and into the convective layer. We say that \textit{under}-
and \textit{over-penetration} occurs when test particles originally
within the convective layer are carried beyond the lower and upper
convective layer boundaries, respectively. Conversely, \textit{bottom}-
and \textit{top-penetration} occurs when test particles initially
bordering the exterior of the convective layer are carried into the
lower and upper boundaries, respectively.

To track the trajectory of tracer particles, the following procedure
is performed at five stages of the burst ($log$  EGR$_{max}$  =
16, 17, 18, 18.5, and 19). At each EGR level, the entire velocity
field is obtained at each timestep for several convective turnover
times. This time-dependent velocity information is inputted into a
separate trajectory analysis algorithm, which evolves the positions
of massless tracer particles using the forward-Euler method. Linear,
2D interpolation is employed to determine approximate velocities when
positions of particles fall between grid-points on the domain. For
each penetration analysis, the positions of two particles for each
of the 385 zones in the lateral direction are tracked. Thus, the positions
of 770 tracer particles are followed per analysis. Particle trajectories
respect the lateral periodicity of the domain, allowing for continuity
of motion at the right edge if particles fly off the left edge, and
vice versa.

\begin{figure}
\plotone{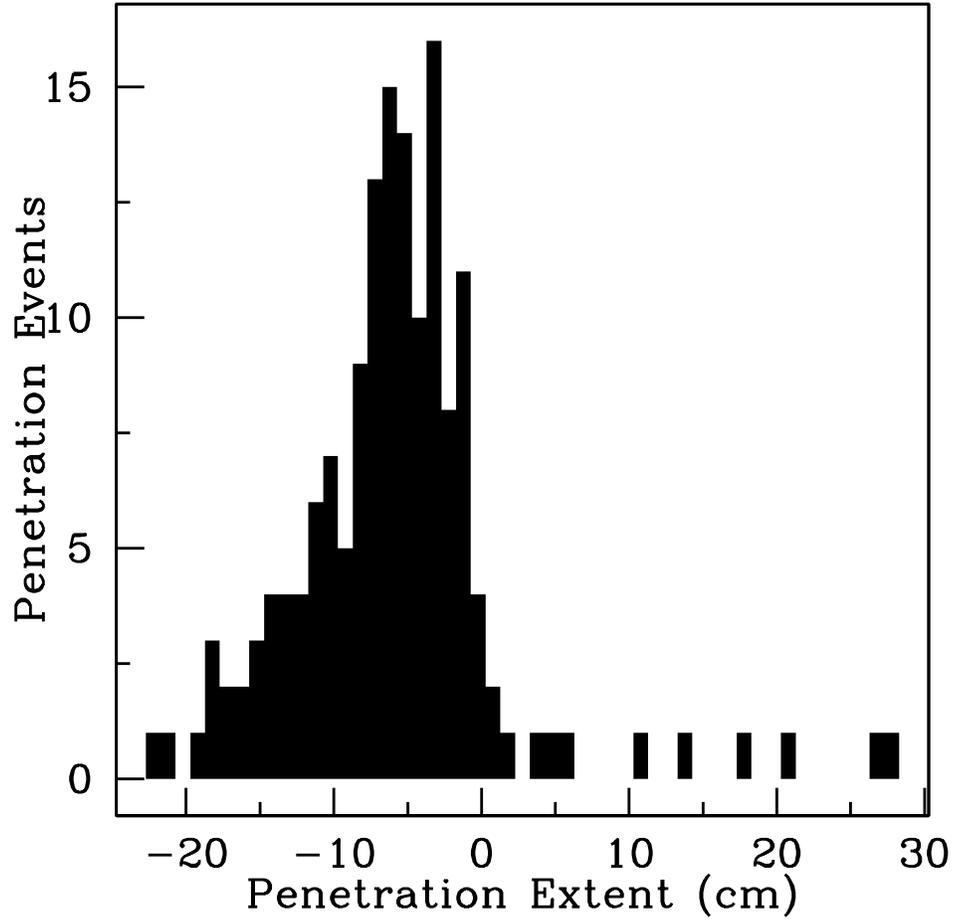}

\caption{{\small Histogram of under- and over-penetration events at $log$
 EGR$_{max}$ = 18.5. Number of events are plotted with respect to
the extent, measured in cm with respect to the positions of the lower
and upper convective layer boundaries. Negative extents indicate under-penetration,
while positive extents, over-penetration.}}

\label{fig:penetration_histogram} 
\end{figure}

Overall, the results of the tracer particle analyses show that when
averaged throughout the burst stages examined, only 10\% of the total
number of particles penetrate the convective layer boundaries, and
when penetration occurs, its extent and duration are limited and temporary.
Virtually no under- and over-penetration occurs when $log$  EGR$_{max}$
$<$ 17, but penetration becomes more frequent at higher EGR levels
when the formal convective layer boundaries fluctuate rapidly on subconvective
timescales. For example, Figure \ref{fig:penetration_histogram} shows
a histogram of under- and over-penetration events at $log$  EGR$_{max}$
= 18.5. The number of penetration events is plotted on the vertical
axis, while the extent of penetration (in cm) with respect to the
time-averaged positions of the lower and upper positions of the convective
layer boundaries is plotted on the horizontal axis. Thus, negative
values of the extent indicate under-penetration, while positive values,
over-penetration. At this particular EGR level, nearly 20\% of the
particles penetrate the convective layer boundaries, mostly under-penetration
events. The majority penetrate less than 10 cm from the lower boundary,
but a few are found as far as 25 cm (0.15 $H_{P}$). Nevertheless,
these penetration events are temporary, and the penetrating particles
are eventually carried back into the convective layer. This effective
trapping of material inside the convective layer may be understood
to result from the dominance of lateral over vertical flows at the
convective layer boundaries (e.g., see Fig. \ref{fig:vw_profile}
and discussion in $\S$3.1). Vertical progress of particles inside
the convective region halts at the boundaries, since the vertical
velocity components diminish by several orders of magnitude there.
Stronger lateral flows deflect the particles until they are eventually
swept back into the convective region. Bottom- and top-penetration
are also very limited. Tracer particles originally above the convective
region are found to more easily fall into it, rather than rise up
from below. Top-penetration depends very sensitively on the initial
positions of the particles. Mixing from beneath the convective layer
(dredge-up) is even more limited.

The maximum extent of penetration is found to be 0.3 $H_{P}$, which
occurs at the bottom boundary at $log$  EGR$_{max}$ = 19, corresponding
to a time when flow speeds are relatively large. Modal analysis of
select lateral slices of the domain in the manner of Herwig et al.
(2006) reveals that near the convective layer boundaries, a gradual
transition from gravity- to convective-modes exists. Thus, convective-modes
are always present where penetration occurs. However, they are also
present when penetration does not occur. Thus, the presence of convective-modes
are necessary but not sufficient for penetration to occur.

\subsection{Effects of convective energy transport}

Here, detailed comparisons of the nuclear flux ($F_{nuc}=\int_{0}^{z_{top}}\rho\dot{s}_{3\alpha}dz$,
where $\dot{s}_{3\alpha}$ is the $3\alpha$ energy generation rate)
between 1D and 2D models are presented. For these comparisons, a 1D
model is evolved with the same initial conditions and parameters as
the 2D model so that any differences can be directly attributed to
the additional mode of energy transfer which convection provides in
2D. The Eddington value of the flux at the surface of a 1.4 M$_{\bigodot}$
neutron star of radius 10$^{6}$ cm is $F_{Edd}=2.5\times10^{25}$
erg s$^{-1}$ cm$^{-2}$, assuming electron scattering, and is used
as a reference value. In principle, the temporal behavior of $F_{nuc}$
during an X-ray burst is directly related to the observed light curve,
since the burst is powered by the thermonuclear event. However, the
present model does not extend up to the actual surface of the neutron
star, and the time evolution of $F_{nuc}$ cannot be rigorously translated
into simulated light curves. Nevertheless, many differences distinguish
the behavior of $F_{nuc}$ in the 1D and 2D models and can be attributable
to the effects of convective dynamics. Thus, convection in neutron
star envelopes arising during a burst may significantly affect what
is actually observed, and simulated light curves from numerical calculations
of this phenomenon need to properly account for convection.

\begin{figure}
\plotone{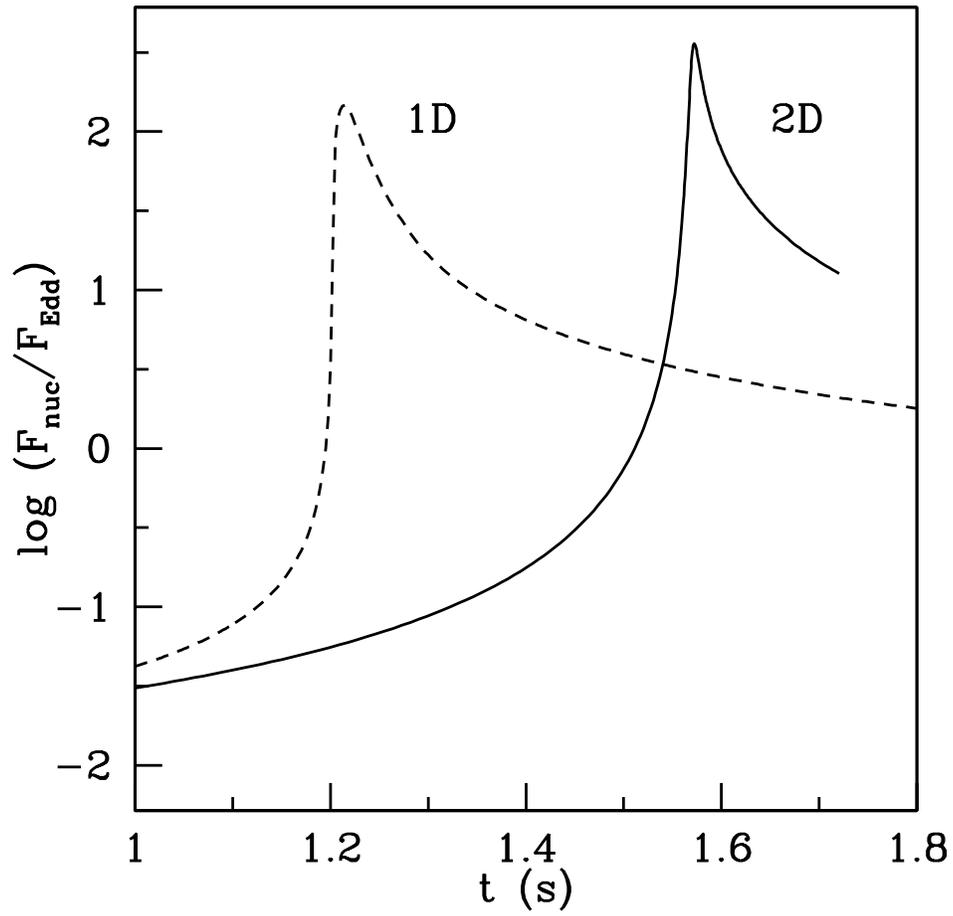}

\caption{{\small \label{fig:1D 2D flux compare}1D, 2D nuclear flux ratio
vs. time. The $log$  of $F_{nuc}/F_{Edd}$ for the 1D (dashed line)
and 2D (solid line) are plotted as a function of physical time (s)
from the start of the calculation ($t$ = 0 s).}}
\end{figure}

In Figure \ref{fig:1D 2D flux compare}, the logarithms  of $F_{nuc}/F_{Edd}$
for both the 1D and 2D models are plotted as functions of time. As
the figure shows, the 1D and 2D temporal evolution of $F_{nuc}$ during
the burst differs significantly. The total time required to reach
burst peak from the beginning of the simulation is delayed 0.36 s
for the 2D model (1.57 s vs. 1.21 s). Likewise, the time required
for $F_{nuc}$ to rise from the Eddington value to the peak value
is over three times greater in 2D than 1D. Similarly, the time required
for $F_{nuc}$ to diminish from $F_{nuc,peak}$ to $e^{-1}F_{nuc,peak}$
in 2D is less than half of what is needed in 1D. These differences
can be understood to be the result of convection, which provides an
additional mode of energy transfer which enhances thermal transport
away from the hottest regions, thus helping to cool and moderate the
nuclear runaway.

The factor of 2.5 increase in the magnitude of the peak nuclear flux
$F_{nuc,peak}$ in the 2D model can also be explained by considering
the dynamic effects of convection and the accompanying advection.
Convection keeps the 2D temperature profile adiabatic. Consequently,
the temperature in the upper part of the 2D domain is several times
greater as compared to the same region in the 1D domain, which exhibits
a very sharp temperature gradient centered on the burning layer. Convection
also directly effects the composition profile, since it thoroughly
mixes fuel from upper regions, where $Y$ is initially greater, to
lower regions, where it is hotter. Consequently, by the time of burst
peak, more fuel has burnt in 2D. The net result of all these differences
is that the EGR in a large part of the the 2D domain exceeds 1D values
by many orders of magnitude at burst peak. For example, at burst peak
at z = 800 cm in the 1D domain, $T_{1D}$ = 2 $\times$ 10$^{8}$
K, $Y_{1D}$ = 0.98, and $log$  EGR$_{max,1D}$ = 11; while at the
same height in the 2D domain, $T_{2D}$ = 5 $\times$10$^{8}$ K,
$Y_{2D}$ = 0.60, and $log$  EGR$_{max,2D}$ = 18. Thus, the greater
$F_{nuc,peak}$ in 2D compared to 1D can be understood to be the result
of convective and advective effects which exclusively occur in the
2D model. Taken together, these results indicate convection significantly
affects energy transport during the burst, resulting in significant
differences in the temporal behavior of $F_{nuc}$ and, presumably,
the actual light curve of the burst.

\section{Code verification}

\begin{figure}
\plotone{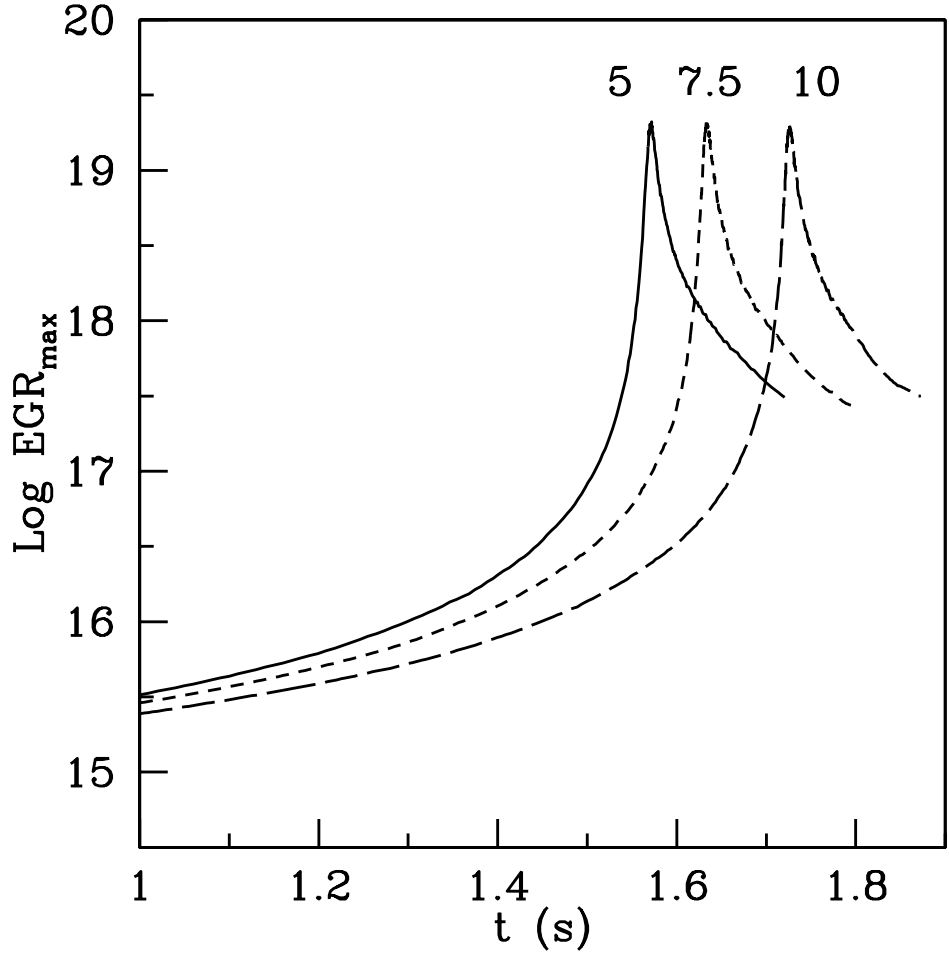}

\caption{{\small \label{fig:validation}Zone-size refinement study. The $log$
 of EGR$_{max}$ for 5, 7.5, and 10 cm/zone models are plotted as
a function of physical time (s) from the start of the calculation
($t$ = 0 s). }}
\end{figure}

Extensive verification studies were performed to demonstrate convergence
of qualitative and quantitative results of the LMNA model. The model
underwent rigorous refinement testing by evolving through complete
burst sequences using different spatial resolutions (5, 7.5, 10 cm/zone),
temporal resolutions (CFL = 0.5, 0.5/2, 0.5/4), and domain sizes (386$\times$200,
386$\times$205). (The purpose of the small change in domain height
in the 386 $\times$ 205 model was to test the sensitivity of the
solution to the upper boundary's placement. The domain height cannot
be greatly extended without encountering limiting factors associated
with lower density regions, such as the severe timestep restrictions
required by thermal diffusion.) At four EGR levels leading up to the
peak of the burst ($log$  EGR$_{max}$  = 16, 17, 18, 19), the different
models were compared according to the characteristics of key features
in the flow fields and how they evolved, the time-evolution of diagnostic
thermodynamic and dynamical quantities (such as EGR$_{max}$, $T_{max}$,
$F_{nuc,max}$, $\vel_{max}$, and $\vel_{ave}$), vertical profiles
of velocity correlations, and thermodynamic gradients $\grad$, $\grad_{ad}$,
and $\grad_{L}$. As an example, Figure \ref{fig:validation} shows
$log$  EGR$_{max}$ as a function of time for three models of 5,
7.5, and 10 cm/zone resolution. The three models attained EGR$_{peak}$
values which agreed to within 3\%. The times required to attain burst
peak ($t_{peak}$) for the 5, 7.5 and 10 cm/zone models are 1.57,
1.63, and 1.73 s respectively, which show a tendency toward convergence
with increasing resolution. (The decrease in trend of $t_{peak}$
can be understood to be the result of a corresponding decrease in
numerical dissipation at greater spatial resolutions.) Similarly,
other diagnostic quantities which were examined agree to within 5\%
of the results of the 5 cm/zone model. Moreover, all of the qualitative
features of the convective dynamics which developed in this model
are reproducible under the spatial, temporal, and domain refinements
which were examined.

\section{Discussion}

The major accomplishments and findings of this project are summarized
as follows: (1) The low Mach number approximation has been developed,
verified, and implemented to study astrophysical deflagrations where
large vertical pressure variations exist and the Mach number $M$
is small. (2) When applied to subadiabatic initial conditions representing
the pre-burst peak stage of a Type I X-ray burst, a vertically expanding
convective layer of Bénard-like cells naturally develops, and the
vertical extent of the larger cells matches that of the convective
layer. The convective layer expands to two pressure scale heights
during the burst progression. (3) Even at their maximum values, convective
flow speeds are substantially subsonic ($M_{peak}<0.15$), while the
deviation of the pressure from the hydrostatic base state is always
at most of order $M^{2}$. (4) As the convective layer expands, fuel
is naturally mixed into the convective layer, and mixing within the
layer is very efficient. However, at the convective layer boundaries,
less efficient mixing results in significant composition gradients,
such that $\grad$ more closely follows $\grad_{L}$ there. Penetration
on convective-timescales is limited and temporary. (5) Both sub- and
superadiabaticity are found within the convective layer, but it is
slightly superadiabatic on average. (6) Convection significantly affects
energy transport.

In the present results, convection develops naturally as a consequence
of superadiabatic gradients arising from heat inputted into the system
by nuclear burning in a bursting layer. No model for convection is
assumed; indeed, we have not been able to establish agreement with
the predictions of mixing-length theory. Throughout the burst, the
average values of the actual gradient $\grad$ are best described
as generally between $\grad_{ad}$ and $\grad_{L}$, while the instantaneous
values are closer to $\grad_{L}$. Moreover, some penetration occurs
at the convective layer boundaries, where on average, $\grad$ clearly
deviates from $\grad_{ad}$ to conform better to $\grad_{L}$. Whether
the Schwarzschild or Ledoux criteria is satisfied in regions where
composition gradients exist in massive stars is still an open question
in astrophysics. Semiconvection, a relatively slow mixing caused by
composition gradients, is poorly understood, and models which examine
the Schwarzschild vs. Ledoux gradients yield conflicting results (Merryfield
1995; Canuto 2000). Moreover, Canuto (2000) demonstrates the Schwarzschild
criteria necessarily implies convective overshooting, and the Ledoux
criteria also necessarily implies overshooting if convection is non-local.
The current model includes all the key elements of convective and
semiconvective processes, and it provides an opportunity to further
examine this important issue.

Several limitations to our current model should be noted. (1) The
contribution of energy generation $\dot{s}_{12,\alpha}$ due to subsequent
$\alpha$ captures on $_{6}^{12}C$ is neglected in the current computation.
Calculations show $\dot{s}_{12,\alpha}$ becomes important post-burst
peak due to the rise in ash concentration $Z$ as a result of burning,
and its inclusion is expected to increase the duration and magnitude
of the simulated burst. (2) At more advanced stages of the burst ($log$
 EGR$_{max}$ $\geq$ 19), the upper convective boundary has reached
the upper domain boundary. Comparison with a model which has an extended
height shows that dynamical results at these stages may be subject
to upper boundary influences. However, these differences were relatively
minor and did not affect the qualitative behavior described here.
Moreover, because the current model does not include the true surface
of the star, observable light curves can not be rigorously calculated.
(3) While most of the computational domain remains degenerate throughout
the burst sequence, calculations show that degeneracy decreases as
the burst progresses, and by the end of the calculation, it begins
to be lifted in the upper third of the domain where the densities
are smallest. Thus, expansion effects may become important at more
advanced stages of the burst, and the current assumption that the
base state is time-independent may need to be relaxed to better model
the dynamics which arise at these later times. (4) The current model
neglects rotation and magnetism. We note that these effects can be
incorporated within the Low Mach number formalism, and we expect to
address them in future work. Nevertheless, the results presented here
describe reasonable qualitative behavior of the flow field as a burst
progresses. The substantially subsonic flows arising during the burst,
the convective layer filling up with larger cells of roughly the same
vertical extent, and the better agreement of $\grad$ with $\grad_{L}$
near the convective layer boundaries are likely to describe the behavior
when these additional effects are included.

Presently, the LMNA method is a powerful computational tool and has
been successfully applied to routinely simulate 2D Type I X-ray burst
deflagrations, a problem which has thus far been intractable with
other methodologies. Continuing to develop the algorithm will be a
vital aspect of future work. Computationally, enhancements may include
relaxing the time-independence of the hydrostatic base state, extending
the model to 3D, implementing different coordinate systems, incorporating
adaptive gridding techniques, and improving the input physics, such
as incorporating rotation, more complete nuclear burning networks,
and a sub-grid turbulence model. The LMNA model is well-suited to
routinely model astrophysical deflagrations which occur during Type
I X-ray bursts, the pre-ejection stage of classical novae, the pre-detonation
stage of supernovae, and the hydrodynamics and burning in the cores
of main sequence stars. The LMNA method represents a useful tool which
enables the routine investigation of a wide variety of interesting
and important astrophysical questions.

\section{Acknowledgments}

We appreciate constructive comments on the draft of this article by
M. Zingale and the anonymous referee. In addition, we also wish to
acknowledge the helpful discussions with P. Ricker and other members
of the FLASH group at the University of Chicago, E. A. Spiegel, M.
M. Mac Low, and A. Heger. This project was supported by NSF grant
DMS-0202485. This work was partially supported by the National Center
for Supercomputing Applications under grant number DMS-020022N and
utilized the Origin2000 Array and the IBM pSeries 690. Other computational
resources during model development, testing, and application were
also generously provided by the FLASH group at the University of Chicago
and the Department of Engineering Sciences and Applied Mathematics
at Northwestern University.

\section{Appendix: The Numerical Procedure for the LMNA}

The numerical procedure for the LMNA is outlined in more detail. A
more complete description of the computational method can be found
in Lin (2006). To enhance readability, the expressions in this Appendix
are given in differential form. The Forward-Euler method is used for
temporal differencing. As will be pointed out below, standard methods
of upwind (advective and first-order convective terms) and central
differencing (diffusion, $gK$, and second-order convective terms)
are used for spatial derivatives. For generality, the equations in
this Appendix are expressed in 3D, where the lateral dimensions are
$x$ and $y$ (or $i$ and $j$), and the vertical dimension is $z$
(or $k$). The LMNA model in the present study is limited to 2D, where
the $x$ ($i$ ) dimension is excluded. All variables have been previously
defined in the text.

At time $t=n$ in zone $i$, $j$, $k$, the variables are: $\rho\power$,
$e\power$, $P\power$, $T\power$, $X\power$, and $\vel\power$.
For each timestep, the hydrodynamic equations are solved in the following
sequence:

1. Solve for the new composition $X\pluspower$ and total reaction
rate $R\pluspower$.\\
The continuity equation for each species $l$ in conservative form
is\[
\ddt{(\rho X_{l})}+\diver{(\rho\vel X_{l})}=\rho R_{l},\]
where $R_{l}$ is the Lagrangian time derivative of species $l$:
$R_{l}\equiv\matder{X_{l}}$. We expand and use the continuity equation,
$\ddt{\rho}+\diver(\rho\vel)=0$, to find the non-conservative form
of the composition equation:

\[
\ddt{X_{l}}=-\vel\cdot\grad{X_{l}}+R_{l}\]
We operator split this step into two operations: (1) advect by upwinding
to a half step, $t=n+\frac{1}{2}$, neglecting the burning rate $R_{l}$;
(2) use $X_{l(i,j,k)}^{(n+\frac{1}{2})},\rho\power,T\power$ in the
$3\alpha$ energy generation routine to obtain the updated composition,
$X_{l(i,j,k)}^{(n+1)}$ and $R_{l(i,j,k)}^{(n+1)}$, where

\[
R_{l}^{(n+1)}\left(X_{l}^{(n+\frac{1}{2})},\rho^{(n)},T^{(n)}\right)=\frac{X_{l(i,j,k)}^{(n+1)}-X_{l(i,j,k)}^{(n+\frac{1}{2})}}{\triangle t}.\]

2. Solve for the new temperature $T\pluspower$. \\
It can be shown that 

\[
\rho\matder{e}-\frac{P}{\rho}\matder{\rho}=\rho c_{p}\frac{DT}{Dt}+-\delta\frac{DP}{Dt}+\sum_{l}\rho\frac{\partial H}{\partial X_{l}}R_{l},\]
where $e$ is the internal energy and $\delta\equiv-\left(\frac{\partial ln\rho}{\partial lnT}\right)_{P}$.
Note that for an ideal gas, $\delta=+1$, but for a general gas, $\delta$
must be explicitly evaluated using partial derivatives from the equation
of state. In the LMNA, the base hydrostatic pressure is assumed to
be constant in time. Then, equating the right-hand-side to $Q+\diver{\kappa\grad T}$
and algebraically manipulating, we obtain the temperature equation:

\[
\ddt{T}=-\vel\cdot\grad T+\frac{1}{c_{p}}\left(\dot{s}-\frac{\delta}{\rho}w\rho_{h}g+\frac{1}{\rho}\diver\kappa\nabla T-\sum_{l}\frac{\partial H}{\partial X_{l}}R_{l}\right),\]
where hydrostatic equilibrium $\grad P_{h}=-\rho_{h}\gvec$ of the
base state is employed.

To update $T$, the temperature equation is operator split into separate
steps. The advection term is upwinded, while the thermal diffusion
term is calculated as two first-order central differences over one
zone: i) to determine the negative of the thermal flux $-F=\kappa\nabla T$
at the edges of a computational zone, and ii) to evaluate $\diver(-F)$
at the zone center. For the thermal diffusion term, $\kappa(\rho,T,X_{l})$,
the radiative thermal conductivity, is $\kappa=\frac{4}{3}\frac{acT^{3}}{\kappa_{o}\rho}$,
where $\kappa_{o}(\rho,T,X_{l})$ is the radiative opacity, calculated
from analytical expressions as referenced in Iben (1975), Christy
(1966), and Weaver et al. (1978).

3. Solve for the new density $\rho\pluspower$. \\
Having updated $X\pluspower$ and $T\pluspower$, and assuming $P\pluspower=P\power$
the equation of state $EOS(T,P_{h},X_{l})$ is used to update density
$\rho\pluspower$ and internal energy $e\pluspower$.

\[
\rho\pluspower\Leftarrow EOS(T,P_{h},X_{l}),\]

\[
e\pluspower\Leftarrow EOS(T,P_{h},X_{l}).\]

4. Solve for $\phi=P^{'}-gK$ using an elliptic equation, where $K=\int_{z_{top}}^{z}\rhoprime(x,y,z^{'})dz^{'}$.\\
To derive the elliptic equation, we rewrite the momentum equation
in terms of $\phi$:

\[
\grad\phi=-\rho\matder{\vel}-\left\{ \dd{(gK)}{x}\ihat+\dd{(gK)}{y}\jhat\right\} .\]
Taking the divergence, we obtain an elliptic equation:

\[
\grad^{2}\phi=-\diver\left(\rho\matder{\vel}\right)-\left\{ \ddd{(gK)}{x}+\ddd{(gK)}{y}\right\} .\]
Using the definition of the material derivative $\matder{}=\ddt{}+\vel\cdot\grad{}$
and the continuity equation, ($\matder{\rho}=-\rho\diver\vel$), $\rho\matder{\vel}$
can be expressed as:

\[
\rho\matder{\vel}=\ddt{(\rho\vel)}+\diver({\rho\vel\vel})\]
\\
Applying the divergence, and again using the continuity equation,
$\diver(\rho\matder{\vel})$ can finally be expressed as:

\[
\diver\left(\rho\matder{\vel}\right)=-\ddd{\rho}{t}+\diver\left\{ \diver(\rho\vel\vel)\right\} .\]
Thus, the Laplacian of $\phi$ becomes:

\[
\grad^{2}\phi=\ddd{\rho}{t}-\diver\left\{ \diver(\rho\vel\vel)\right\} -\left\{ \ddd{(gK)}{x}+\ddd{(gK)}{y}\right\} .\]
This elliptic equation for $\phi$ is solved using FISHPAK, a package
of subprograms for the solution of separable, elliptic partial differential
equations developed by Adams, Swarztrauber, and Sweet (1988). 

We next consider the detailed evaluation of each of the terms on the
right-hand-side of the elliptic equation:

a.) Using the continuity equation, the time derivative of $\ddd{\rho}{t}$
is expressed as

\[
\ddd{\rho}{t}=\frac{1}{\Delta t}\left[\left(\ddt{\rho}\right)^{n+1}+\diver(\rho\vel)^{n}\right],\]
where at time-level $n+1$, $\left(\ddt{\rho}\right)^{n+1}$ is calculated
analytically from an expression obtained by evaluating the partial
derivatives of the internal energy, and at time-level $n$, $\diver(\rho\vel)^{n}$
is centrally differenced over two zones. This procedure for calculating
$\ddd{\rho}{t}$ was found to be required for numerical stability. 

b.) By considering its parts and then taking the divergence, a relatively
simple and symmetric expression is obtained for $\diver\left\{ \diver(\rho\vel\vel)\right\} $,
which include the convective terms:

\begin{eqnarray*}
\diver\left\{ \diver(\rho\vel\vel)\right\}  & = & \ddd{(\rho u^{2})}{x}+\ddd{(\rho v^{2})}{y}+\ddd{(\rho w^{2})}{z}\\
 &  & +\dd{}{x}\dd{(\rho uv)}{y}+\dd{}{x}\dd{(\rho uw)}{z}\\
 &  & +\dd{}{y}\dd{(\rho vu)}{x}+\dd{}{y}\dd{(\rho vw)}{z}\\
 &  & +\dd{}{z}\dd{(\rho wu)}{x}+\dd{}{z}\dd{(\rho wv)}{y}.\end{eqnarray*}
Each second-order convective term is calculated as two first-order
derivatives: i) upwinding is used first to obtain the terms in $\diver({\rho\vel\vel})$,
and ii) these terms are then centrally differenced over one zone to
obtain the final expressions in $\diver\left\{ \diver(\rho\vel\vel)\right\} $.

c.) Derivatives involving $K$ are centrally differenced. 

5. Solve for the new velocity $\vel\pluspower$. \\
From the momentum equation, 

\[
\ddt{(\rho\vel)}=-\diver{(\rho\vel\vel)}-\grad\phi-\left\{ \dd{(gK)}{x}\ihat+\dd{(gK)}{y}\jhat\right\} .\]
The convective terms which had been previously calculated are applied
again here. Derivatives involving $K$ are centrally differenced.
Because the domain utilizes a staggered-grid system, the velocity
components are determined and stored at zone edges. To determine a
zone's central values of the velocity components, the appropriate
component's values at the zone edges are spatially averaged

6. Finally, the timestep is complete, and all quantities are updated
to $t=n+1$: \\
$\rho\pluspower$, $e\pluspower$, $P_{(i,j,k)}^{\prime(n+1)}$, $T\pluspower$,
$X\pluspower$, $\vec{v}\pluspower$ .

\clearpage

\clearpage

\clearpage

\clearpage 

\clearpage

\clearpage

\begin{figure}
\plotone{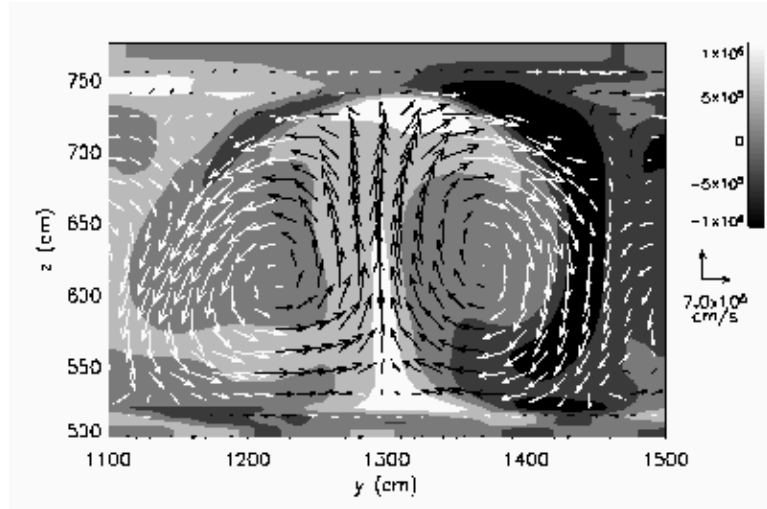}

\caption{{\small Bénard cell at $log$  EGR$_{max}$ = 17. Velocity vectors
are superimposed over contours of} $\Delta T${\small , the difference
of a zone's temperature compared to the lateral mean. Vertical and
lateral coordinates given with respect to the lower left corner ($y=0,$
$z=0$ ) of the domain. Black velocity vectors are used when $w>0$;
white, when $w<0$. Upflows are associated with $\Delta T>0$, while
downflows, $\Delta T<0$. Every 3rd velocity vector is plotted.}}

\label{fig:benard_cell_17} 
\end{figure}

\clearpage

\clearpage

\clearpage

\clearpage

\clearpage

\clearpage

\clearpage 
\end{document}